\documentclass[
reprint,
amsmath,amssymb,
aps,
pre,
floatfix,
]{revtex4-2}

\usepackage{xcolor}
\usepackage{multirow}
\usepackage{graphicx}
\usepackage{subcaption}
\usepackage{hyperref}
\usepackage{xurl}

\begin{document}

\title{Statistical Landscape Exploration and Weak-Field Defects in Selective Oscillator Ising Machines}

\author{Ömer Önder}
 \email{omer.onder@bilkent.edu.tr}
\affiliation{%
 Department of Physics, Bilkent University, Ankara 06800, T{\" u}rkiye
}%
\author{Aydın Cem Keser}%
 \email{aydin.keser@bilkent.edu.tr}

\affiliation{%
 Department of Physics, Bilkent University, Ankara 06800, T{\" u}rkiye
}%

\date{\today}

\begin{abstract}
Oscillator Ising machines (OIMs) provide an analog dynamical approach to
combinatorial optimization, but comparatively little is known about the
statistical structure of the ensembles produced by their dynamics. Using
the Sherrington--Kirkpatrick (SK) spin-glass model as a representative benchmark,
we introduce a selective oscillator scheme, OIM+, in which successive
annealing epochs are combined with energy-based elimination and
repopulation of replicas. We characterize the resulting states through
overlap distributions, Hamming distances, hierarchical clustering, and
local-field statistics, and show that OIM+ explores structured low-energy
regions of the frustrated landscape while retaining a nontrivial ensemble
of configurations despite repeated selection.
We identify a sparse population of weak-local-field residual defects as a
characteristic limitation of the selective oscillator dynamics. This
failure mode contrasts with the familiar difficulty of local-update
annealing, where strongly correlated clusters can resist reorganization
through sequences of unfavorable single-spin moves. A short
Metropolis--Hastings refinement preferentially corrects the weak-field
defects left by OIM+, supporting a picture in which the oscillator dynamics performs the collective exploration while the remaining corrections are predominantly local. The weak fields arise mainly from cancellation among
competing interaction terms rather than from uniformly weak couplings,
identifying frustration-induced local balance as their microscopic origin.
The same qualitative behavior persists for an antiferromagnetically biased
SK ensemble, where the hybrid method reaches a pooled mean energy comparable
to that of a substantially longer pure Metropolis anneal. The complete
protocol also achieves competitive performance on standard G-set Max-Cut
benchmarks. These results provide a physical picture of how selective
oscillator dynamics explores frustrated energy landscapes and suggest a
natural division of labor between collective analog search and targeted
local refinement.
\end{abstract}

\maketitle



\section{Introduction}
\label{sec:intro}
Originally introduced as a minimal description of ferromagnetism, the Ising
model has since become a paradigmatic framework for studying collective
phenomena, phase transitions, disorder, and frustration across a wide range
of physical systems \cite{SK1,SK2,Nishimori2001,BinderYoung1986}. Beyond its role in condensed matter physics, the Ising Hamiltonian has acquired renewed importance as a generic representation of combinatorial optimization problems. A broad class of NP-hard and NP-complete problems, including graph partitioning, satisfiability, scheduling, portfolio optimization, and variants of the traveling-salesman problem, can be reformulated as the search for low-energy configurations of an Ising system or its equivalent quadratic unconstrained binary optimization (QUBO) representation \cite{Barahona1982,Lucas_2014}. Consequently, efficient methods for exploring rugged Ising energy landscapes and locating their low-energy regions are of interest not only for understanding equilibrium and nonequilibrium statistical physics, but also for solving computationally challenging optimization problems.

A variety of numerical approaches have been developed to search for low-energy Ising configurations, ranging from exact branch-and-bound techniques and message-passing methods to Monte Carlo sampling, simulated annealing, parallel tempering, and more recent machine-learning-based approaches. Among these, simulated annealing remains one of the most widely used and conceptually
influential methods \cite{Simu_Ann}. By combining stochastic sampling with a gradual reduction of an effective temperature, simulated annealing exploits thermal fluctuations to escape local minima and explore rugged energy landscapes. Its success has established a close connection between statistical-physics concepts and combinatorial optimization, making annealing-based methods a natural reference point  for assessing alternative optimization paradigms.

Despite its broad applicability, conventional single-spin local-update Monte Carlo annealing can experience increasingly slow relaxation in frustrated systems with rugged energy landscapes. In dense spin glasses, for example, large-scale reorganization of correlated spin clusters may require a sequence of energetically unfavorable intermediate moves, leading to long equilibration times and trapping in metastable states \cite{BinderYoung1986,Young1997,MPV1987}. Parallel tempering and nonlocal cluster updates can mitigate some of these limitations, including on probabilistic-computing hardware \cite{Chowdhury2025ProbabilisticComputers}. Such single-spin limitations motivate optimization strategies capable of exploiting collective dynamics that reorganize correlated degrees of freedom beyond local spin flips.

The growing importance of Ising optimization has stimulated interest in specialized hardware and physics-inspired algorithms that exploit the dynamics of physical systems to search for low-energy states. Beyond conventional digital implementations, a broad family of Ising machines has emerged, including coherent Ising machines, digital annealers, simulated bifurcation machines, quantum annealers, and oscillator-based architectures \cite{machines,Delacour2026ProbabilisticReview,machines1,machines2,Goto2021}. The motivation behind these approaches is that optimization can be embedded
directly into the evolution of a physical system, allowing collective physical
dynamics to explore the energy landscape and perform part of the computational
task \cite{OIM}; large programmable quantum annealers provide one prominent
example of this principle \cite{King2023QuantumCritical}.

Such architectures are particularly attractive for large-scale optimization
problems, where the computational demands of conventional approaches motivate
the development of specialized physical computing platforms
\cite{machines,Neumann}. Frustration and local ruggedness can further complicate annealing-based exploration ~\cite{King2019Ruggedness}.

Among these approaches, oscillator Ising machines (OIMs) have emerged as a
promising fully classical paradigm in which binary spin variables are encoded
in the phases of coupled nonlinear oscillators \cite{OIM,Neumann}. OIMs can
be implemented on a variety of electronic and photonic platforms and have
demonstrated competitive performance on benchmark optimization problems such
as Max-Cut and related graph-partitioning tasks \cite{Energy3,Energy4}.
Unlike single-spin Monte Carlo annealing, which explores the energy landscape through
discrete local updates, OIMs exploit continuous collective synchronization
dynamics that can simultaneously reorganize many effective spins. Related
work on coherent Ising machines has interpreted optimization as geometric
landscape annealing, in which changing control parameters progressively
reshape the effective landscape encountered by the dynamics
\cite{Yamamura2024Landscape}. This provides a complementary picture of how
collective analog dynamics can guide the search through an evolving energy
landscape.

Because the search dynamics can be carried out directly by a physical
oscillator network, OIMs may ultimately enable low-latency and energy-efficient
optimization in dedicated hardware. At the same time, their collective
continuous dynamics raises a distinct statistical question: whether the
ensembles generated by OIMs explore frustrated energy landscapes differently
from those produced by conventional local-update algorithms.

Recent work on physical Ising machines has increasingly emphasized that
useful information need not be limited to the single lowest-energy
configuration found in a run. Quantum-annealing studies have examined the
organization of sampled low-energy configurations into distinct states,
clusters, and basins \cite{Zhang2024Cyclic,Zhang2025Complexity}, while
coherent Ising machines have been used to sample ground and low-energy spin
configurations \cite{Ng2022CIMSampling}. Noise-injected analog Ising
machines have also been experimentally demonstrated as statistical samplers
of Boltzmann distributions \cite{Bohm2022StatisticalSampling}. Together,
these developments motivate characterizing physical Ising machines through
the ensembles generated by their dynamics, in addition to their best
attained energies.

For oscillator Ising machines, studies have emphasized optimization
performance, benchmark energies, hardware implementations, and scaling
characteristics \cite{OIM,machines,Energy3,Energy4}, while recent work has
also configured OIMs as stochastic p-bit engines
\cite{Ekanayake2026PBitOIM}. Less is known specifically about the
configuration-space structure of low-energy ensembles generated by selective
oscillator dynamics and the residual local stability after that selective
evolution. Recent work has identified spin freezing
induced by second-harmonic injection as an important dynamical limitation
of oscillator Ising machines \cite{Farasat2026}. Here, we examine these
questions for selective oscillator annealing by resolving both the
configuration-space structure of the resulting ensemble and the local
stability of the residual defects.

These questions are particularly relevant for frustrated systems and spin
glasses, where low-energy-state organization, overlap distributions, and
state diversity contain information beyond the best energy attained
\cite{MPV1987,Rammal1986,Wales2003}. They are also relevant to optimization
settings in which access to several high-quality solutions is useful:
near-optimal configurations can differ in structural properties,
robustness, or compatibility with additional constraints. Related
quantum-annealing work has emphasized the value of accessing multiple deep
configurations rather than only a single minimum \cite{ZhangKamenev2025NN}, and
similar objectives arise in quality-diversity and related search paradigms
\cite{QD,ExtractQD,PortfolioQD}. Understanding how OIM dynamics populates such low-energy ensembles is therefore important both for optimization and for elucidating how analog dynamical systems explore complex energy landscapes.

The aim of this work is therefore not simply to benchmark an oscillator
optimizer against a conventional annealing method. Rather, we ask how
selective oscillator dynamics populates a frustrated low-energy landscape,
what configurational structure survives the selection process, and which
degrees of freedom remain locally unresolved once the collective oscillator
search has ended.

To address these questions, we use the fully connected SK model as a
controlled and well-understood setting in which to resolve the statistical
mechanisms of the oscillator dynamics. We study a selective oscillator
protocol, which we denote OIM+. For each fixed SK coupling realization, we
evolve a population of stochastic oscillator replicas: all replicas share the
same coupling matrix, but begin from distinct random phase configurations and
follow independent noise histories. Successive epochs of collective oscillator
annealing are interleaved with energy-based selection and repopulation. After
each epoch, the higher-energy replicas are removed and the surviving low-energy
configurations are replicated before oscillator evolution resumes.

Population-level selection and repopulation have precedents in population
annealing \cite{Wang2015GroundStates}. OIM+ combines this broad
population-based idea with nonequilibrium oscillator dynamics and a direct
energy-ranked selection rule. In the calculations presented here, the
resulting dynamics explores structured low-energy regions of the spin-glass
landscape while retaining a diverse ensemble of configurations. This suggests
that selective oscillator dynamics performs the main collective exploration
of low-energy configuration space rather than merely searching for a single
minimum-energy state.

A short Metropolis--Hastings refinement, used as a diagnostic probe of the
residual defects, produces what we denote defect-healed OIM+ (DH--OIM+).
Comparison of OIM+ and DH--OIM+ reveals a distinct residual failure mode:
the corrected spins are concentrated at weak or vanishing local fields,
arising predominantly from cancellation among competing interactions rather
than from uniformly weak couplings.

\begin{figure*}[t]
\centering
\includegraphics[width=\linewidth]{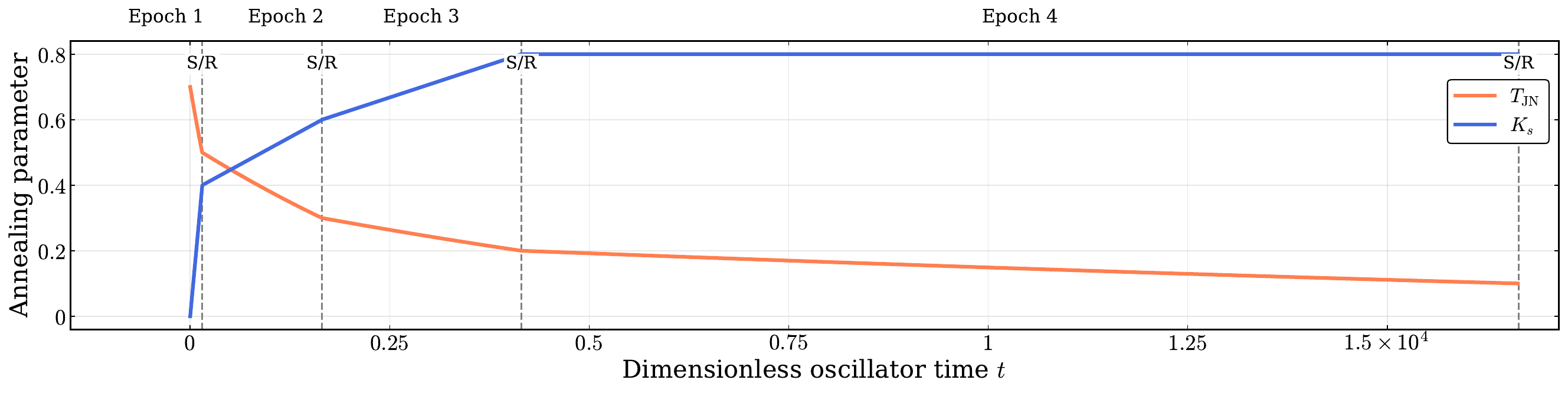}%
\caption{
Selective multi-epoch annealing schedule used in OIM+.
The protocol consists of four successive oscillator-annealing epochs with
progressively lower effective noise temperature $T_{\rm JN}$ and stronger
SHIL binarization strength $K_s$.
Within each epoch, the noise amplitude $K_n$ is reduced linearly, producing
the corresponding decrease in $T_{\rm JN}=K_n^2/2$, while $K_s$ is increased
linearly except in the final epoch, where $K_s$ is held fixed.
At the end of each oscillator epoch, an energy-based selection step eliminates
the highest-energy fraction $f_{\rm cull}=2/3$ of the replica population,
and the surviving configurations are replicated to restore the original
population size.
After the first three selection--repopulation steps, the replicated states
undergo the subsequent oscillator epoch, while the fourth selection defines
the final OIM+ population.
The combination of continuous oscillator dynamics and post-epoch
selection--repopulation defines the OIM+ protocol.
The horizontal axis shows the dimensionless oscillator time $t$ defined in
Sec.~\ref{sec:anneal}.
For the complete set of simulation parameters, see
Tables~\ref{tab:annealing_schedule} and~\ref{tab:simulation_parameters}.}
\label{fig:annealing_schedule}
\end{figure*}

For the antiferromagnetically biased ensemble, the selectively evolved
population followed by the short defect-healing sweep reaches a pooled mean
energy comparable to that of the independent pure-MH baseline, while using
more than an order of magnitude fewer sequential Metropolis proposals in its
refinement stage. We additionally test the complete DH--OIM+ protocol on
heterogeneous G-set Max-Cut benchmarks that need not be fully connected,
where it achieves competitive performance
\cite{Helmberg2000,Goudet2024}.

These results provide a physical picture of how selection-assisted oscillator
dynamics explores frustrated energy landscapes, how the annealing mechanism
shapes the statistical structure of the resulting ensemble, and where
residual limitations remain.

{The remainder of this paper is organized as follows.
Section~II introduces the selective oscillator dynamics, annealing protocol,
and statistical diagnostics. Section~\ref{sec:results} characterizes the optimization
behavior, ensemble structure, residual weak-field defects, and their
dependence on antiferromagnetic bias. Section~\ref{sec:benchmarks} evaluates the approach on
G-set Max-Cut benchmarks. Sections~\ref{sec:disc} and \ref{sec:conc} present the discussion and
conclusions, respectively.}

\section{Model and Selective Oscillator Annealing}
\label{sec:model}

The primary objective of this work is to understand how selective oscillator
dynamics explores frustrated Ising landscapes, what statistical structure it
generates, and which failure modes remain after annealing. To this end, we
focus primarily on the Sherrington--Kirkpatrick (SK) spin glass, which
provides a canonical model of a densely connected frustrated optimization
landscape and a natural testbed for overlap statistics, state diversity, and
hierarchical organization. In this setting, we study a selective oscillator
protocol, denoted OIM+, in which collective phase dynamics is organized into
successive annealing epochs, each followed by energy-based selection and
repopulation of the replica population.

Here, a replica denotes one stochastic realization of the system for a fixed
SK disorder realization. In OIM+, each replica is an oscillator-network
trajectory: all replicas associated with a given SK instance share the same
coupling matrix, but are initialized with distinct random phase configurations
and evolve under independent stochastic noise histories. At a selection--repopulation step, descendants inherit the selected
oscillator state and subsequently evolve under independent stochastic forcing.
Further details of the disorder generation, replica randomization, and
reproducibility procedure are given in Appendix~\ref{app:randomization}.

To assess whether the resulting optimization mechanism extends beyond the
mean-field spin-glass setting, we additionally consider a collection of
standard G-set Max-Cut benchmark instances.

Unless otherwise stated, the simulations use the epoch-dependent annealing
schedule illustrated in Fig.~\ref{fig:annealing_schedule} and detailed in
Table~\ref{tab:annealing_schedule}, with the global parameters summarized in
Table~\ref{tab:simulation_parameters}.

\subsection{Sherrington--Kirkpatrick spin glass}

The SK model is defined by the Ising Hamiltonian

\begin{equation}
E(\mathbf{s})=
-\sum_{i<j} J_{ij}s_i s_j,
\qquad
s_i\in \lbrace-1,+1\rbrace,
\label{eq:SK}
\end{equation}
where every spin interacts with every other spin through quenched random couplings $J_{ij}$. The model provides a paradigmatic example of a frustrated mean-field spin glass and exhibits a rugged energy landscape containing a large number of metastable states and competing low-energy basins \cite{SK1,SK2,BinderYoung1986,MPV1987}.

\begin{table*}[t]
\caption{
Four-epoch oscillator-annealing schedule used in OIM+.
$N_{\rm int}$ denotes the number of numerical integration steps,
$\Delta t$ the corresponding integration time step, and
$t_{\rm epoch}=N_{\rm int}\Delta t$ the accumulated integration time
during each epoch.
}
\label{tab:annealing_schedule}
\begin{ruledtabular}
\begin{tabular}{cccccc}
Epoch &
$N_{\rm int}$ &
$\Delta t$ &
$T_{\rm JN}^{(0)}\!\rightarrow T_{\rm JN}^{(1)}$ &
$K_s^{(0)}\!\rightarrow K_s^{(1)}$ &
$t_{\rm epoch}$ \\
\hline
1 & $15\times10^4$ & $10^{-3}$ &
$0.7\rightarrow0.5$ &
$0\rightarrow0.4$ &
$150$ \\

2 & $15\times10^4$ & $10^{-2}$ &
$0.5\rightarrow0.3$ &
$0.4\rightarrow0.6$ &
$1500$ \\

3 & $25\times10^4$ & $10^{-2}$ &
$0.3\rightarrow0.2$ &
$0.6\rightarrow0.8$ &
$2500$ \\

4 & $25\times10^4$ & $5\times10^{-2}$ &
$0.2\rightarrow0.1$ &
$0.8\rightarrow0.8$ &
$12500$ \\
\hline
Total & & & & & $16650$ \\
\end{tabular}
\end{ruledtabular}
\end{table*}

Throughout most of this work, the couplings are independently drawn from a Gaussian distribution,
\begin{equation}
J_{ij}
\sim
\mathcal N\left(\frac{J_0}{N},\frac{\sigma_J^2}{2N}\right),
\end{equation}
with symmetric interactions, $J_{ij}=J_{ji}$, and no self-couplings, $J_{ii}=0$.
The factor of $1/2$ in the variance reflects the matrix construction used here: the initially sampled coupling matrix is symmetrized according to $J\rightarrow(J+J^{T})/2$, which halves the off-diagonal variance. We choose the coupling scale as the unit of energy, setting \(\sigma_J=1\) throughout, while the mean coupling \(J_0\) is varied to control the interaction bias. We primarily consider the conventional SK model with $J_0=0$, corresponding to a symmetric distribution of ferromagnetic and antiferromagnetic couplings. To examine the behavior under an antiferromagnetic bias, in Sec.~\ref{sec:antiferro} we additionally consider $J_0=-4$, which shifts the coupling distribution toward antiferromagnetic interactions while preserving the dense random connectivity of the underlying graph.
Further details of the disorder construction, replica initialization,
stochastic-noise generation, and random-seed hierarchy are given in
Appendix~\ref{app:randomization}.

The SK model is particularly well suited to the present study for two
reasons. First, its dense connectivity resembles optimization problems that
naturally give rise to dense QUBO or Ising formulations, while also providing
a canonical setting for studying frustration in a fully connected system.
Although all-to-all coupling is more demanding for some hardware
architectures than sparse connectivity, dense or fully connected interactions
are not intrinsically outside the scope of Ising-machine implementations:
all-to-all connectivity has been demonstrated in coherent Ising machines
\cite{machines2} and in coupled-oscillator CMOS architectures
\cite{Lo2023,Budrikis2024}. Second, the model possesses a rich glassy landscape with a
nontrivial organization of low-energy states that can be characterized
through overlap distributions, hierarchical clustering, and local-field
statistics. We therefore use the SK model primarily as a well-understood
mechanistic testbed for resolving how selective oscillator dynamics explores
and populates frustrated energy landscapes, while the G-set benchmarks
provide a complementary test on more heterogeneous graph structures.

\subsection{Oscillator Ising machine}

The underlying analog dynamics of OIM+ is provided by an oscillator Ising machine (OIM) consisting of a network of $N$ coupled phase oscillators \cite{OIM,Neumann}. Each oscillator is associated with a vertex of the underlying Ising graph, and its phase $\phi_i(t)$ evolves according to
\begin{equation}
\dot{\phi}_i =
-
\frac{1}{2}\sum_{j\neq i}
J_{ij}
\sin(\phi_i-\phi_j)
-K_s(t)\sin(2\phi_i)
+K_n(t)\xi_i(t),
\label{eq:oim_dynamics}
\end{equation}
where $K_s(t)$ denotes the second-harmonic injection-locking (SHIL) strength, which progressively promotes phase binarization, $K_n(t)$ controls the stochastic forcing, and $\xi_i(t)$ is a Gaussian white-noise process satisfying
\begin{equation}
\langle \xi_i(t)\rangle =0,
\qquad
\langle \xi_i(t)\xi_j(t')\rangle=
\delta_{ij}\delta(t-t').
\end{equation}
The annealing protocol is implemented through the time-dependent noise amplitude $K_n(t)$. Equivalently, the stochastic forcing may be characterized by an
equivalent Johnson--Nyquist noise temperature
\begin{equation}
\label{eq:Tjn}
T_{\rm JN}(t)
=
\frac{K_n^2(t)}{2},
\end{equation}
which determines the strength of the oscillator noise.

The parameters $K_s$ and $K_n$ are varied during the annealing process according to a prescribed schedule described below in Sec.~\ref{sec:anneal}.
The stochastic oscillator equations in Eq.~\eqref{eq:oim_dynamics} are
integrated numerically using the Euler--Maruyama method.
The integration time step $\Delta t$ is allowed to vary between annealing
epochs, as specified in Table~\ref{tab:annealing_schedule}.

For fixed $K_s$ and in the absence of noise, the dynamics relaxes toward minima of the Lyapunov function
\begin{equation}
\mathcal{E}(\boldsymbol{\phi})
=
-\frac{1}{2}\sum_{i<j}
J_{ij}
\cos(\phi_i-\phi_j)
-\frac{K_s}{2}
\sum_i
\cos(2\phi_i),
\label{eq:lyapunov}
\end{equation}
where $K_s=K_s(t)$ is varied during the annealing schedule.
As $K_s$ increases, the SHIL term progressively confines the oscillator phases to the binary states
$\phi_i\in\lbrace 0,\pi\rbrace$. Identifying these phase states with Ising spins $s_i=\pm1$, the Lyapunov function reduces on the binary manifold to the Ising Hamiltonian up to a scaling and an additive constant,
\begin{equation}
\mathcal{E}(\boldsymbol{\phi})
=
-\frac{1}{2}\sum_{i<j}J_{ij}s_i s_j
-\frac{N}{2}K_s ,
\end{equation}
so that minimizing the oscillator energy is equivalent to minimizing the target Ising energy.

\begin{table}[t]
\caption{Default parameters of the selective OIM+ protocol used for the
SK-model simulations.}
\label{tab:simulation_parameters}
\begin{ruledtabular}
\begin{tabular}{lc}
Parameter & Value \\
\hline
System size & $N=200$ \\
Replica population & $N_{\rm rep}=402$ \\
Coupling scale & $\sigma_J=1$ \\
Number of oscillator epochs & $4$ \\
Selection culling fraction & $f_{\rm cull}=2/3$ \\
Total cumulative integration time & $t_{\rm tot}=16650$ \\
\end{tabular}
\end{ruledtabular}
\end{table}

To reduce trapping in metastable configurations, we employ a multi-stage
annealing protocol with post-epoch energy selection and repopulation,
which we denote OIM+. The protocol consists of four successive
oscillator-annealing epochs, with progressively reduced noise and stronger
phase binarization. After each epoch, higher-energy replicas are removed
and the surviving configurations are replicated to restore the population
size. The detailed annealing and selection schedules are described in the
following subsection.

\subsection{Multi-epoch annealing protocol}
\label{sec:anneal}

The oscillator dynamics is evolved using a multi-stage annealing procedure designed to balance exploration of the continuous phase landscape with progressive binarization of the oscillator states. The OIM+ protocol consists of four successive annealing epochs, separated by energy-based selection and repopulation of the replica ensemble.
Within each annealing epoch, the effective noise amplitude
$K_{n}$ and the SHIL strength $K_s$ are varied according to
prescribed linear schedules. If $p\in[0,1]$ denotes the normalized progress through an epoch, the noise
amplitude is varied linearly according to
\begin{equation}
K_n(p)
=
K_n^{(0)}(1-p)
+
K_n^{(1)}p,
\end{equation}
where the endpoint values are set by the corresponding Johnson--Nyquist
temperatures,
\begin{equation}
K_n^{(0,1)}
=
\sqrt{2T_{\rm JN}^{(0,1)}}.
\end{equation}
The effective temperature during the epoch then follows Eq.~\eqref{eq:Tjn}.
The SHIL strength is varied according to
\begin{equation}
K_s(p)
=
K_s^{(0)}(1-p)
+
K_s^{(1)}p.
\end{equation}
The endpoint values $T_{\rm JN}^{(0,1)}$ and $K_s^{(0,1)}$ differ
between epochs and are listed in Table~\ref{tab:annealing_schedule}.
In the final epoch, $K_s^{(0)}=K_s^{(1)}$, so that the SHIL strength
remains fixed while the effective temperature continues to decrease.
Because $\Delta t$ differs between annealing epochs, the number of numerical
integration steps alone does not represent the duration of the simulated
oscillator evolution. We therefore use the cumulative integration time
\begin{equation}
t = \sum_e n_e \Delta t_e ,
\label{eq:cumulative_time}
\end{equation}
where $n_e$ denotes the number of integration steps taken in epoch $e$ and
$\Delta t_e$ is the corresponding integration time step. For the four-epoch
schedule used here, the total cumulative integration time is
$t_{\rm tot}=16650$. The numerical values used in each annealing epoch are listed in
Table~\ref{tab:annealing_schedule}.

The integration step $\Delta t$ is a numerical discretization parameter and
should not be interpreted as an elementary operation of a physical OIM.
In an analog implementation, the oscillator phases evolve continuously and
in parallel, with the physical time scale determined by the oscillator
frequency, coupling strength, injection-locking dynamics, and device response
time. For reference, demonstrated electronic oscillator Ising machines operate
on nanosecond time scales: a 1968-node CMOS ring-oscillator implementation
operating near $1\,\mathrm{GHz}$ reached its ground-state configuration within
fewer than 50 oscillator cycles, corresponding to a relaxation time below
approximately $50\,\mathrm{ns}$ \cite{Energy4}. Recent oscillator
implementations span similar orders of magnitude; for example, a
$9.1\,\mathrm{GHz}$ MEMS oscillator network exhibits a synchronization time of
approximately $90\,\mathrm{ns}$, while a monolithically integrated
optoelectronic Ising machine reports a round-trip time of $1.71\,\mathrm{ns}$
and a spin-evolution time of approximately $150\,\mathrm{ns}$
\cite{Deng2024,Wu2025}. These measurements indicate characteristic device
time scales from sub-nanoseconds to nanoseconds and collective relaxation
times typically of order $10$--$10^2\,\mathrm{ns}$ in fast current
implementations, although the precise mapping depends on the hardware
architecture. Consequently, the number of OIM integration steps used in a
numerical simulation is not directly comparable to the number of sequential
spin-update proposals in a digital Monte Carlo anneal. We therefore
characterize the simulated OIM evolution by its cumulative integration time
rather than by its integration-step count.

The combination of high initial temperature and weak SHIL provides an
extended exploratory regime in which the oscillator phases remain only
weakly constrained. As annealing proceeds, the decreasing temperature and
increasing SHIL progressively drive the phases toward the binary states.
In the final epoch, the SHIL strength is held fixed while the temperature
continues to decrease.

At the end of each  oscillator epoch, the continuous phase configuration is mapped onto Ising spins according to
\begin{equation}
s_i=\mathrm{sign}[\cos(\phi_i)].
\label{eq:phase_binarization}
\end{equation}
Thus, phases closer to $0$ modulo $2\pi$ are mapped to $s_i=+1$, whereas phases closer to $\pi$ modulo $2\pi$ are mapped to $s_i=-1$. In the measure-zero case $\cos(\phi_i)=0$, we assign $s_i=+1$; this convention has no practical effect on the results. Consequently the replicas
are ranked by their Ising energies from Eq.~\eqref{eq:SK}. An energy-based
selection step is then applied with culling fraction
\begin{equation}
f_{\mathrm{cull}}=\frac{2}{3}.
\end{equation}
The highest-energy two thirds of the population are eliminated, leaving the
lowest-energy third. Two additional copies of each surviving configuration
are then created, restoring the replica population from $134$ to $402$.
The duplicated replicas inherit the selected oscillator state but subsequently
evolve under distinct stochastic noise streams; implementation details are
given in Appendix~\ref{app:randomization}. The initially identical replicas
therefore diverge during the subsequent oscillator epoch.

After the first three epochs, the replicated configurations undergo the next
oscillator epoch. The same selection and
repopulation operation is also applied after the fourth epoch, thereby
defining the final selected OIM+ population prior to the diagnostic
Metropolis refinement. The complete annealing and selection schedule is
shown in Fig.~\ref{fig:annealing_schedule}, while the remaining global
simulation parameters are summarized in
Table~\ref{tab:simulation_parameters}.

\begin{figure*}[t]
    \centering
    \begin{subfigure}[b]{0.49\textwidth}
        \centering
        \includegraphics[width=\linewidth]{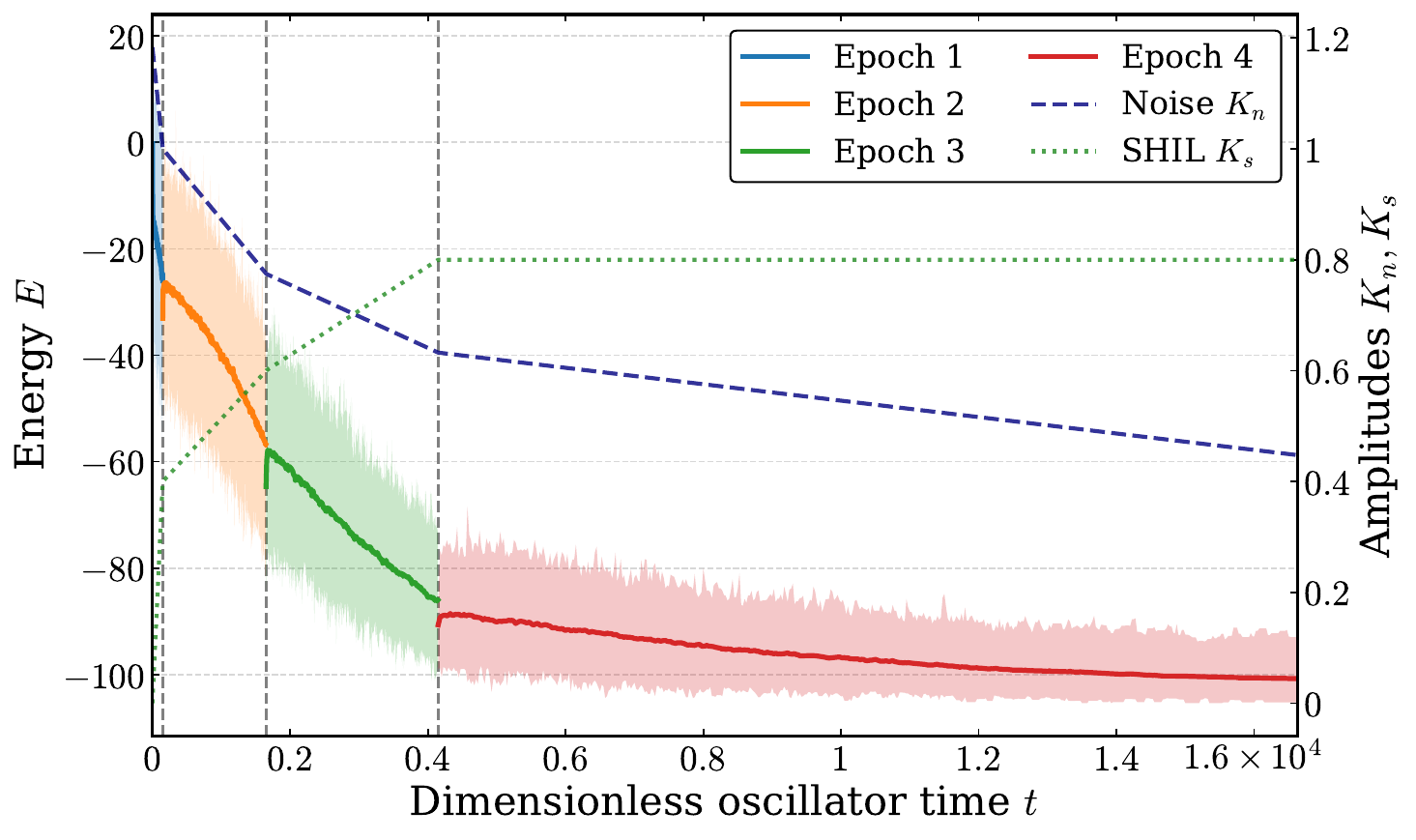}
        \caption{}
        \label{fig:traj_oimp}
    \end{subfigure}
    \hfill
    \begin{subfigure}[b]{0.49\textwidth}
        \centering
        \includegraphics[width=\linewidth]{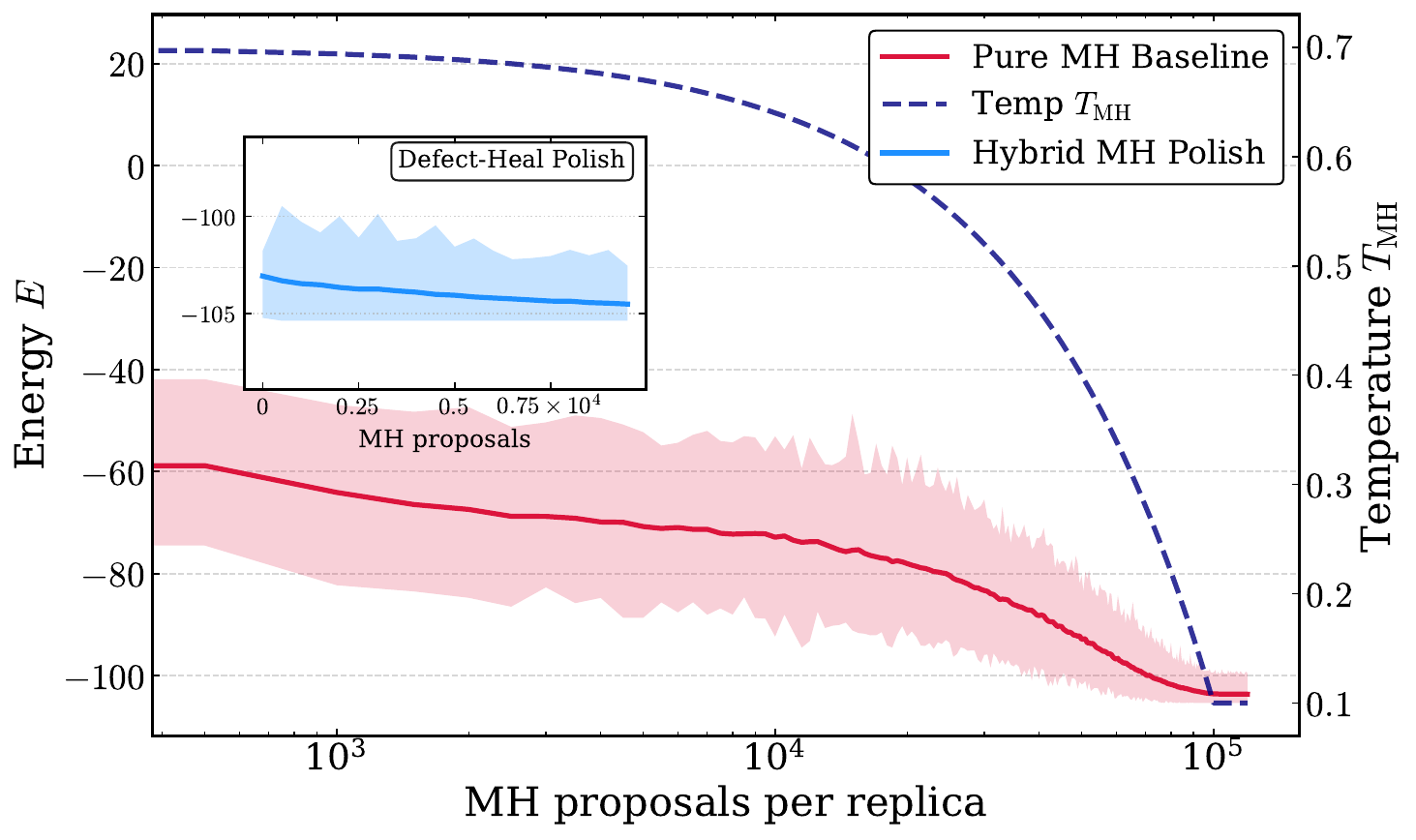}
        \caption{}
        \label{fig:traj_mh}
    \end{subfigure}
\caption{
	Energy evolution for the optimization protocols for a representative
	$J_0=0$ SK disorder realization.
	(a) Evolution of the OIM+ population across the four oscillator-annealing
	epochs as a function of the dimensionless oscillator time $t$.
	The solid curve shows the mean Ising energy over the 402 replicas,
	while the shaded region spans the minimum and maximum energies across the
	population. Vertical dashed lines indicate the boundaries between successive
	oscillator epochs. The dashed and dotted curves show the noise amplitude
	$K_n$ and SHIL strength $K_s$, respectively.
	(b) Energy evolution of the independent pure Metropolis--Hastings baseline
	for the same disorder realization, with the solid curve and shaded region
	again showing the population mean and minimum--maximum range over 402
	replicas. The dashed curve gives the MH temperature $T_{\rm MH}$.
	The inset shows the short $10^4$-proposal diagnostic MH refinement applied
	to the final selected OIM+ population on its own MH-proposal axis.
	The minimum energy sampled during the OIM+ trajectory is approximately
	$-105.37$; the same minimum is subsequently reached during the diagnostic
	MH refinement.
	Thus, for this particular disorder realization, the diagnostic MH refinement
	does not further lower the best energy encountered during the OIM+ evolution,
	although it further lowers the energy of the final selected population and
	modifies the population-level structure analyzed below.
}
\label{fig:energy_evolution}
\end{figure*}

\subsection{Metropolis refinement and pure-MH baseline}
\label{sec:healing}

Metropolis--Hastings (MH) dynamics is used in this work for two distinct
purposes. First, a short MH sweep is applied after OIM+ as a local
refinement and diagnostic probe of the residual defects left by the
selective oscillator dynamics. Second, a substantially longer pure-MH
anneal is used as an independent optimization baseline. The two protocols
use the same single-spin Metropolis update rule but differ in their
initial conditions, annealing schedules, and number of updates.

For the diagnostic refinement, we start from the final selected OIM+
population and map the oscillator phases onto Ising spins according to
Eq.~\eqref{eq:phase_binarization}. A trial move consists of selecting a
spin $s_i$ uniformly at random and proposing the flip
$s_i\rightarrow -s_i$. The corresponding change in the Ising Hamiltonian is
\begin{equation}
\Delta E
=
2s_i\sum_{j\neq i}J_{ij}s_j ,
\label{eq:mh_deltaE}
\end{equation}
and the proposed move is accepted according to the Metropolis--Hastings
rule \cite{MH}, with probability
\begin{equation}
P_{\rm acc}
=
\min\!\left[
1,
\exp\!\left(-\frac{\Delta E}{T_{\rm MH}}\right)
\right].
\label{eq:mh_accept}
\end{equation}
Here $T_{\rm MH}$ denotes the Metropolis acceptance temperature.

With the energy normalization used here, $\mathcal{E}=E/2+\mathrm{const.}$ on the binary manifold, so the corresponding temperature scales are related by
\begin{equation}
T_{\rm JN}= T_{\rm MH}/2.
\end{equation}

The diagnostic refinement consists of $10^4$ single-spin update proposals
per replica, during which $T_{\rm MH}$ is reduced linearly from $0.2$ to
$0.1$. This sweep is deliberately short and is not intended to perform a
second global optimization. Instead, comparison of the configurations
before and after the sweep identifies the spins that remain susceptible to
local correction after OIM+ has completed its collective search. When a
separate label is required, we refer to the resulting post-refinement
ensemble as  defect-healed OIM+ (DH--OIM+).

To provide an independent local-update benchmark, we additionally perform
pure-MH annealing starting from random Ising configurations. Each replica
undergoes $10^5$ single-spin update proposals while the MH temperature is
reduced linearly from $0.7$ to $0.1$. The system is then evolved for a
further $10^4$ settling updates at $T_{\rm MH}=0.1$, followed by a
$10^4$-update measurement interval at the same temperature. During the
measurement interval, the spin orientations are accumulated in time, and
the reported final MH configuration is obtained from the sign of the
time-averaged spin state. Thus, the pure-MH benchmark contains
$1.2\times10^5$ single-spin update proposals per replica in total.
The two Metropolis protocols are summarized in
Table~\ref{tab:mh_protocols}.

\begin{table}[htb]
\caption{
Metropolis--Hastings protocols used for the post-OIM+ diagnostic
refinement and the independent pure-MH benchmark. All update counts
refer to single-spin proposals per replica.
}
\label{tab:mh_protocols}
\begin{ruledtabular}
\begin{tabular}{lcc}
Parameter & Diagnostic refinement & Pure MH \\
\hline
Initial state
& Selected OIM+ state
& Random spins \\

Replica population
& $402$
& $402$ \\

Annealing updates
& $10^4$
& $10^5$ \\

Initial $T_{\rm MH}$
& $0.2$
& $0.7$ \\

Final annealing $T_{\rm MH}$
& $0.1$
& $0.1$ \\

Settling updates
& ---
& $10^4$ \\

Measurement updates
& ---
& $10^4$ \\

Total updates
& $10^4$
& $1.2\times10^5$ \\
\end{tabular}
\end{ruledtabular}
\end{table}

\subsection{Statistical diagnostics}
To characterize the ensembles generated by OIM+, the post-refinement DH--OIM+ states, and conventional Metropolis--Hastings annealing, we employ several complementary diagnostics probing optimization performance, state similarity, hierarchical organization, and local stability. Together, these measures allow us to relate the energetic quality of the obtained solutions to their underlying statistical structure and to identify the residual limitations of the selective oscillator dynamics.

The Ising energy of a spin configuration $\mathbf{s}$ is computed from Eq.~(\ref{eq:SK}) and serves as the primary optimization metric. For oscillator-generated states, this energy is evaluated after phase binarization according to Eq.~(\ref{eq:phase_binarization}); the continuous oscillator Lyapunov energy is not used as the reported optimization metric. In addition to average energies, we examine the full distribution of energies obtained across the replica population in order to assess the energetic quality and spread of the generated ensembles.

\begin{figure*}[htb!]
    \centering
    \begin{subfigure}[b]{0.49\textwidth}
        \centering
        \includegraphics[width=\textwidth]{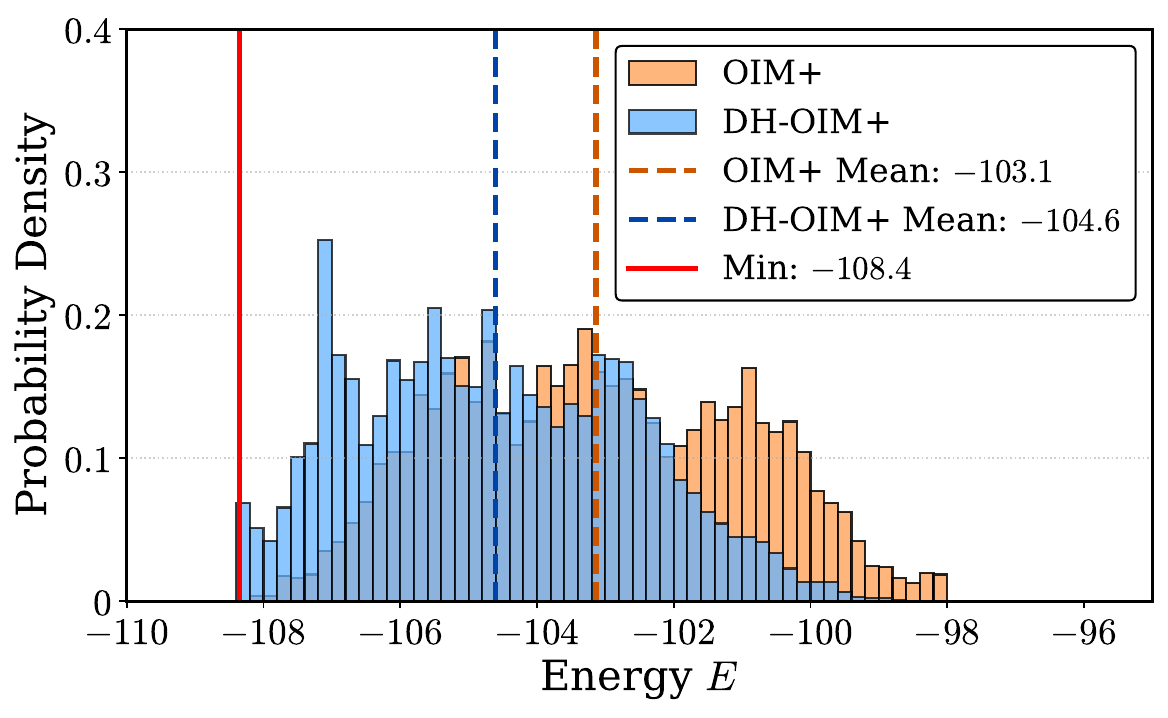}
        \caption{}
    \end{subfigure}
    \hfill
    \begin{subfigure}[b]{0.49\textwidth}
        \centering
        \includegraphics[width=\textwidth]{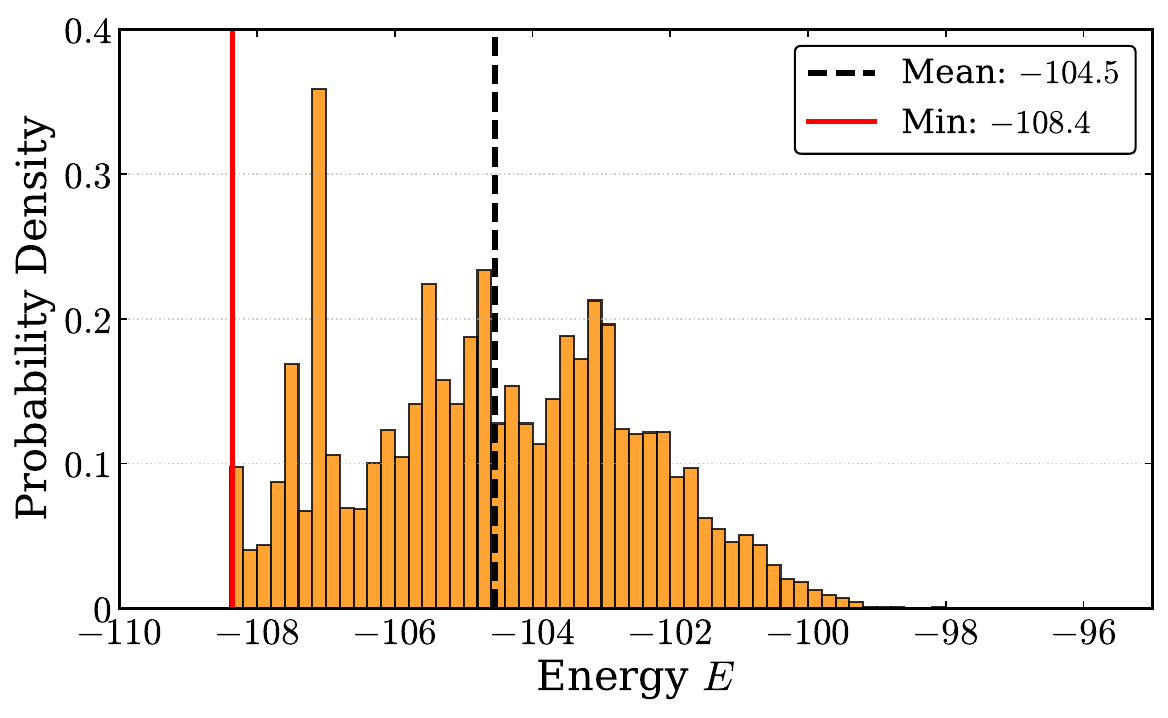}
        \caption{}
    \end{subfigure}
    \caption{Distribution of final Ising energies for the $J_0=0$ SK ensemble,
pooled over 30 independent disorder realizations with 402 replicas per
realization ($12\,060$ configurations for each distribution).
(a) Final energies of the selected OIM+ population before and after the
short diagnostic Metropolis--Hastings refinement. The refinement shifts
the pooled mean energy from $-103.1$ for OIM+ to $-104.6$ for
DH--OIM+ and reduces the high-energy tail.
(b) Final-energy distribution obtained from the independent pure-MH
baseline evaluated over the same disorder ensemble, with pooled mean
energy $-104.5$. The minimum energies observed over the full pooled
ensembles are $E_{\min}^{\mathrm{DH-OIM+}}=-108.4$ and
$E_{\min}^{\mathrm{MH}}=-108.4$.}
    \label{fig:energy_histogram}
\end{figure*}

To quantify similarity between configurations, we use the spin overlap
\begin{equation}
q_{\alpha\beta}=
\frac{1}{N}
\sum_{i=1}^{N}
s_i^{(\alpha)}
s_i^{(\beta)},
\end{equation}
where $\alpha$ and $\beta$ label two replica configurations. The overlap distribution provides a standard characterization of spin-glass structure and forms the basis of the Parisi overlap formalism \cite{MPV1987}. Large positive values of $q_{\alpha\beta}$ indicate similar states, while values near zero correspond to weakly correlated configurations. Large negative values indicate configurations related by extensive spin inversion; in the zero-field SK model, the global transformation $\mathbf{s}\rightarrow-\mathbf{s}$ reflects the underlying $\mathbb{Z}_2$ symmetry.

As is standard for the Sherrington--Kirkpatrick spin glass, overlap
statistics are averaged over independent realizations of the quenched
disorder rather than inferred from a single coupling realization
\cite{MPV1987}. For a given disorder realization $\mathcal{J}_a$,
let $P_{\mathcal{J},a}(q)$ denote the corresponding overlap distribution.
We report the disorder-averaged distribution
\begin{equation}
P(q)
=
\frac{1}{N_{\text{seeds}}}
\sum_{a=1}^{N_{\text{seeds}}}
P_{\mathcal{J},a}(q),
\end{equation}
where $N_{\text{seeds}}=30$ independent disorder realizations are used
unless otherwise stated.

Closely related to the overlap is the normalized Hamming distance
\begin{equation}
d_H(\alpha,\beta)
=
\frac{1-q_{\alpha\beta}}{2},
\end{equation}
which gives the fraction of spins that differ between configurations
$\alpha$ and $\beta$. Because the zero-field SK Hamiltonian is invariant under
the global transformation $\mathbf{s}\rightarrow-\mathbf{s}$, configurations
related only by a global spin inversion should be identified in the
distance-based analysis. We therefore define the
$\mathbb{Z}_2$-quotiented Hamming ($\mathbb{Z}_2$-Hamming for short) distance
\begin{equation}
d_{\mathbb Z_2}(\alpha,\beta)
=
\frac{1-|q_{\alpha\beta}|}{2}
=
\min\!\left[
d_H(\alpha,\beta),
1-d_H(\alpha,\beta)
\right].
\end{equation}
Under this definition, two globally inverted configurations have zero
distance. All pairwise-distance, hierarchical-clustering, cophenetic,
intra-cluster, and inter-cluster quantities reported below are computed
using $d_{\mathbb Z_2}$.

The signed overlap $q_{\alpha\beta}$ is retained when constructing the
disorder-averaged overlap distributions $P(q)$. Thus, the overlap
distributions preserve the distinction between positive and negative
overlaps, whereas the clustering analysis characterizes configuration
space modulo the global $\mathbb Z_2$ symmetry.

To visualize the organization of low-energy states, we perform hierarchical clustering using the pairwise $\mathbb{Z}_2$-Hamming distance matrix and represent the resulting hierarchy by dendrograms, tree diagrams in which the merge height indicates the distance at which configurations or clusters are joined~\cite{Zhang2025Complexity}. These diagnostics provide a compact representation of how configurations are organized in configuration space and allow us to compare the degree of hierarchical clustering produced by the different annealing dynamics.

For two configurations, the cophenetic distance is the linkage height at
which their branches first merge in the dendrogram. We quantify how faithfully
the hierarchy represents the underlying geometry using the cophenetic
correlation coefficient, defined as the Pearson correlation between the
pairwise $\mathbb{Z}_2$-Hamming distances and their cophenetic distances. It therefore quantifies how faithfully the dendrogram represents the underlying pairwise-distance structure.

To investigate the local origin of the residual corrections made by the
diagnostic MH refinement, we analyze the local effective field~\cite{Palmer1979} acting on
each spin,
\begin{equation}
h_i
=
\sum_{j\neq i} J_{ij}s_j .
\end{equation}
The magnitude $|h_i|$ measures the energetic bias acting on spin $i$.
Spins experiencing weak local fields are only weakly stabilized by the
surrounding Ising configuration and therefore have small energetic
penalties for reversal. As shown below, the spins corrected by the
diagnostic MH refinement are strongly concentrated in this weak-field
population, motivating their interpretation as residual weak-field defects
of OIM+.

Finally, to identify the subset of spins corrected by the diagnostic Metropolis refinement, we define the corrected-spin set
\begin{equation}
\mathcal{H}
=
\left\lbrace
i :
s_i^{\rm OIM+}
\neq
\tilde{s}_i^{\rm DH-OIM+}
\right\rbrace,
\end{equation}
where $s_i^{\rm OIM+}$ and $\tilde{s}_i^{\rm DH-OIM+}$ denote the spin orientations before and after the Metropolis refinement stage, respectively. Before evaluating $\mathcal H$, the post-refinement configuration is aligned with the OIM+ configuration under the global $\mathbb Z_2$ symmetry $\mathbf{s}\rightarrow-\mathbf{s}$, so that a global inversion is not counted as a set of local corrections. Comparing the local-field statistics of $\mathcal{H}$ with those of the unchanged spin population provides a direct probe of the residual local defects corrected by the diagnostic refinement.

\begin{figure*}[t]
    \centering
    \begin{subfigure}[b]{0.49\textwidth}
        \centering
        \includegraphics[width=\textwidth]{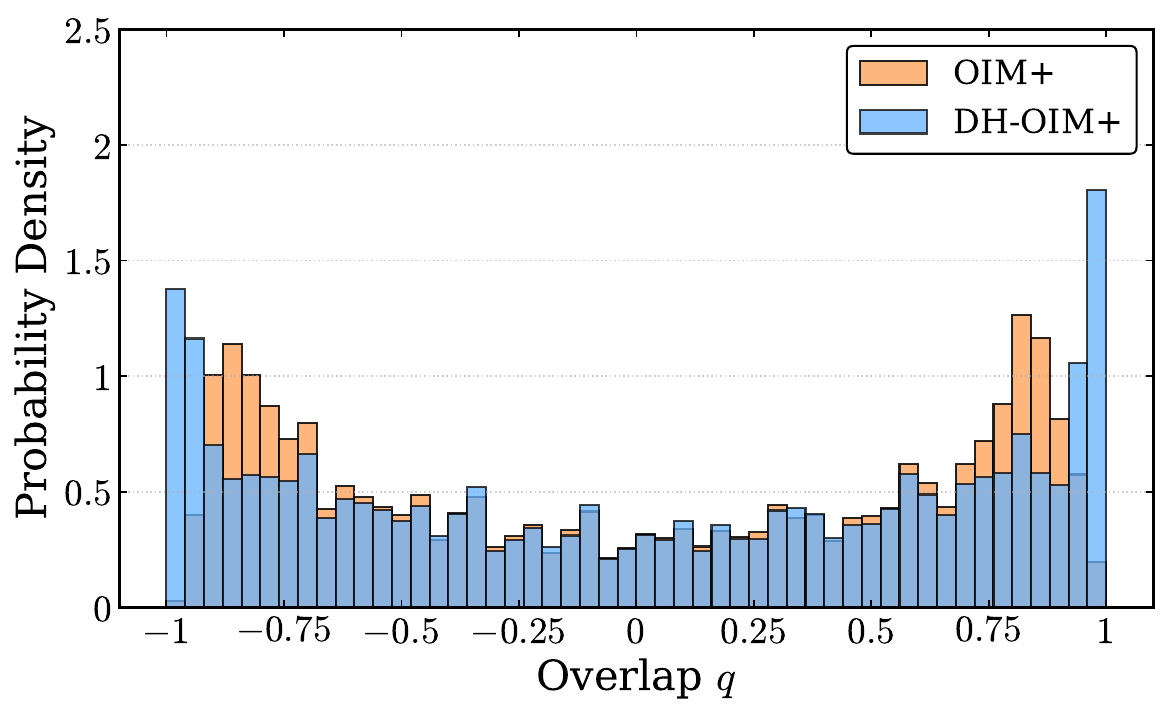}
        \caption{}
    \end{subfigure}
    \vspace{0.1cm}
    \begin{subfigure}[b]{0.49\textwidth}
        \centering
        \includegraphics[width=\textwidth]{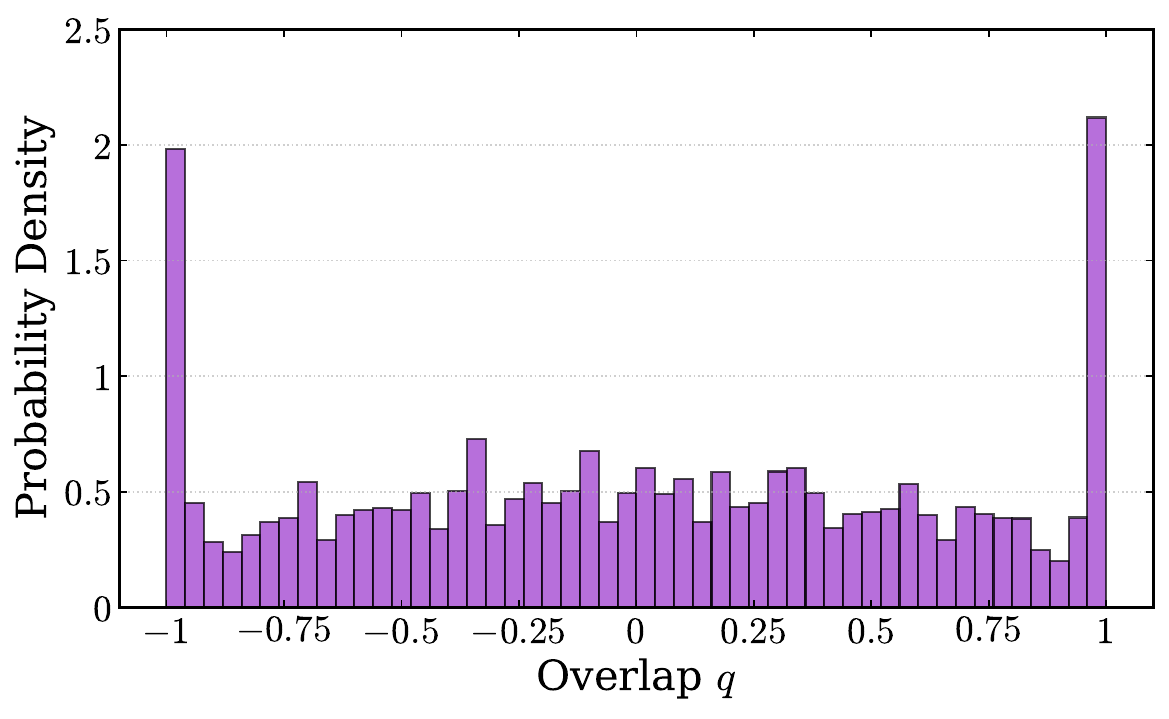}
        \caption{}
    \end{subfigure}
    \caption{
Disorder-averaged overlap distributions $P(q)$ for the $J_0=0$
SK ensemble, averaged over 30 independent disorder realizations.
For each realization, overlaps are computed between replica configurations
generated for the same coupling matrix $J$, and the resulting distributions
are subsequently averaged over disorder.
(a) Comparison of the selected OIM+ population before refinement and the
corresponding post-refinement DH--OIM+ population.
(b) Overlap distribution obtained from the independent pure-MH baseline.
All three ensembles retain substantial weight at intermediate overlap,
showing that the low-energy populations contain distinct partially
correlated configurations rather than collapsing exclusively onto
$q=\pm1$ states.
}
\label{fig:qmatrix}
\end{figure*}

\section{Results}
\label{sec:results}

\subsection{Energy relaxation and defect healing}

We begin by examining the evolution of the energy during OIM+ and the
subsequent diagnostic MH refinement. Figure~\ref{fig:energy_evolution}
shows the corresponding energy evolution for one representative $J_0=0$
disorder realization, including the four OIM+ oscillator-annealing epochs,
the subsequent short diagnostic Metropolis refinement, and the independent
pure-MH baseline. The energy decreases rapidly during the oscillator
evolution, with successive epochs producing progressively smaller
improvements. By the end of the fourth oscillator epoch, the OIM+
population has already reached a low-energy region of the landscape, while
the subsequent MH refinement produces only a comparatively modest
additional decrease in the mean energy of the replica population.

For the representative disorder realization shown in
Fig.~\ref{fig:energy_evolution}, the OIM+ trajectory transiently reaches
an energy of approximately $-105.37$ during the oscillator evolution.
After the final selection step, the minimum energy of the selected OIM+
population is approximately $-105.22$, while the subsequent diagnostic
MH refinement lowers it to $-105.37$. The refinement also lowers the mean
energy of the representative population from approximately $-103.1$ to
$-104.6$. The refinement consists of only $10^4$ single-spin proposals
per replica, as summarized in Table~\ref{tab:mh_protocols}. Thus, this
representative realization already shows that OIM+ brings the population
into a low-energy region, while the short Metropolis stage provides
additional local improvement. The ensemble-level diagnostics below
determine more systematically the division of roles between collective
oscillator exploration and residual local correction.

To assess the robustness of this observation, Fig.~\ref{fig:energy_histogram} compares the distributions of final energies obtained from OIM+, the corresponding post-refinement DH--OIM+ states, and the independent pure Metropolis--Hastings baseline. While OIM+ already produces low-energy configurations, the diagnostic MH refinement systematically shifts the ensemble toward lower energies and reduces the incidence of high-energy outliers. The resulting improvement is obtained with only \(10^4\) additional single-spin MH proposals per replica, consistent with the interpretation of the refinement as a local correction to the preceding collective search.

These results motivate two related questions: how are the low-energy
configurations generated by OIM+ organized in configuration space, and which
residual spins remain susceptible to local correction? We address the first
through overlap and clustering diagnostics and the second through local-field
statistics.

\subsection{Statistical structure of OIM ensembles}

The energy-based diagnostics of the previous section indicate that OIM+
already reaches low-energy configurations before the diagnostic
Metropolis refinement. Energy alone, however, does not reveal how these
configurations are organized within the underlying spin-glass landscape.
We therefore examine the statistical structure of the ensembles generated
by OIM+, the corresponding post-refinement DH--OIM+ states, and the
independent pure-MH baseline using overlap distributions, Hamming
distances, and hierarchical clustering.

The final-energy distributions in Fig.~\ref{fig:energy_histogram} contain
pronounced concentrations at several energy values, indicating that many
replicas reach energetically similar configurations. These configurations
do not, however, collapse onto a single spin state. Their pairwise $\mathbb{Z}_2$-Hamming
distances remain finite, showing that comparable energies can correspond
to distinct configurations within the spin-glass landscape.
This behavior
is consistent with the presence of multiple distinct, near-degenerate
low-energy configurations.

Figure~\ref{fig:qmatrix} compares the disorder-averaged overlap
distributions obtained from OIM+, DH--OIM+, and pure MH. All three
protocols retain substantial weight at intermediate values of $q$ rather
than concentrating exclusively near $q=\pm1$, showing that the low-energy
ensembles contain distinct, partially correlated configurations. The OIM+
and DH--OIM+ distributions remain qualitatively similar, indicating that
the short diagnostic refinement does not strongly reorganize the overlap
structure. Nevertheless, differences between the OIM-based ensembles and
pure MH are visible: OIM+ and DH--OIM+ retain substantial intermediate-overlap
weight, whereas pure MH exhibits more pronounced peaks near $q=\pm1$ and
a comparatively flatter intermediate region. Thus, the optimization
protocol affects not only the energies reached but also the configurational
statistics of the resulting ensembles. The organization underlying these
pairwise similarities is examined more directly through hierarchical
clustering of the $\mathbb{Z}_2$-Hamming distances and the quantitative
measures summarized in Table~\ref{tab:distance_metrics}.
\begin{figure}[t]
    \centering
    \begin{subfigure}[b]{\linewidth}
        \centering
        \includegraphics[width=\textwidth]{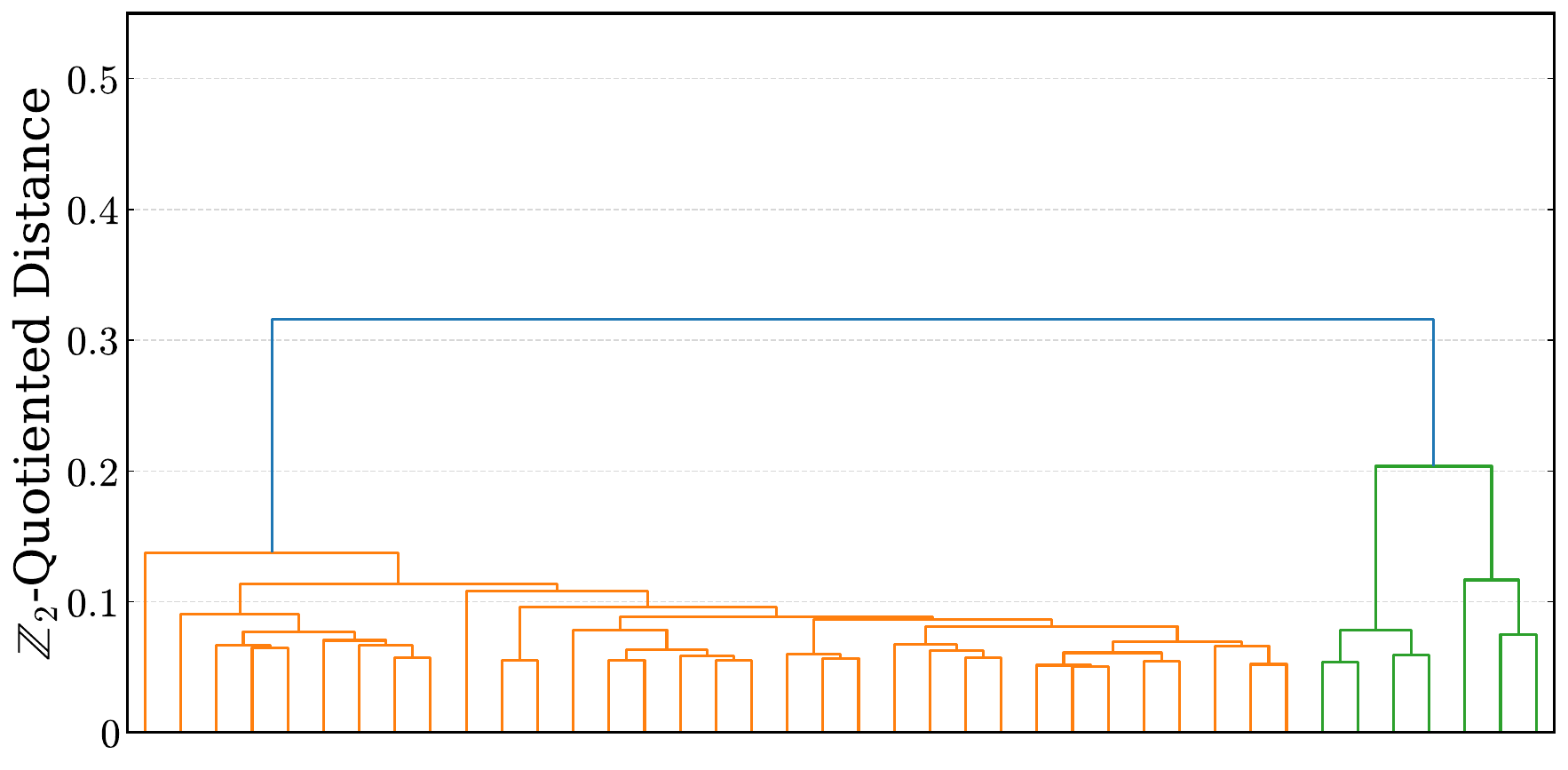}
        \caption{}
    \end{subfigure}
    \vspace{-0.1cm}

    \begin{subfigure}[b]{\linewidth}
        \centering
        \includegraphics[width=\textwidth]{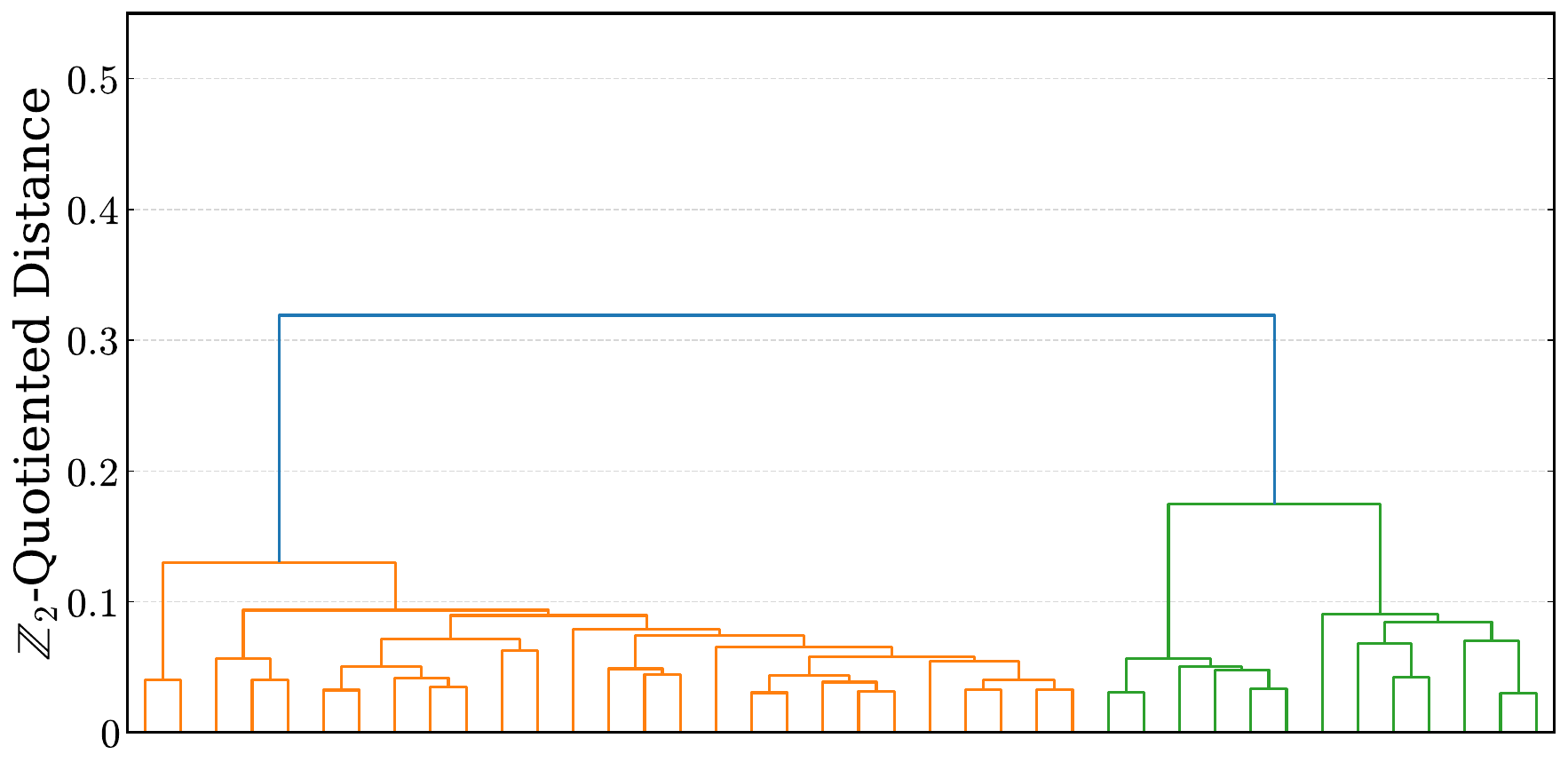}
        \caption{}
    \end{subfigure}
    \vspace{-0.1cm}

    \begin{subfigure}[b]{\linewidth}
        \centering
        \includegraphics[width=\textwidth]{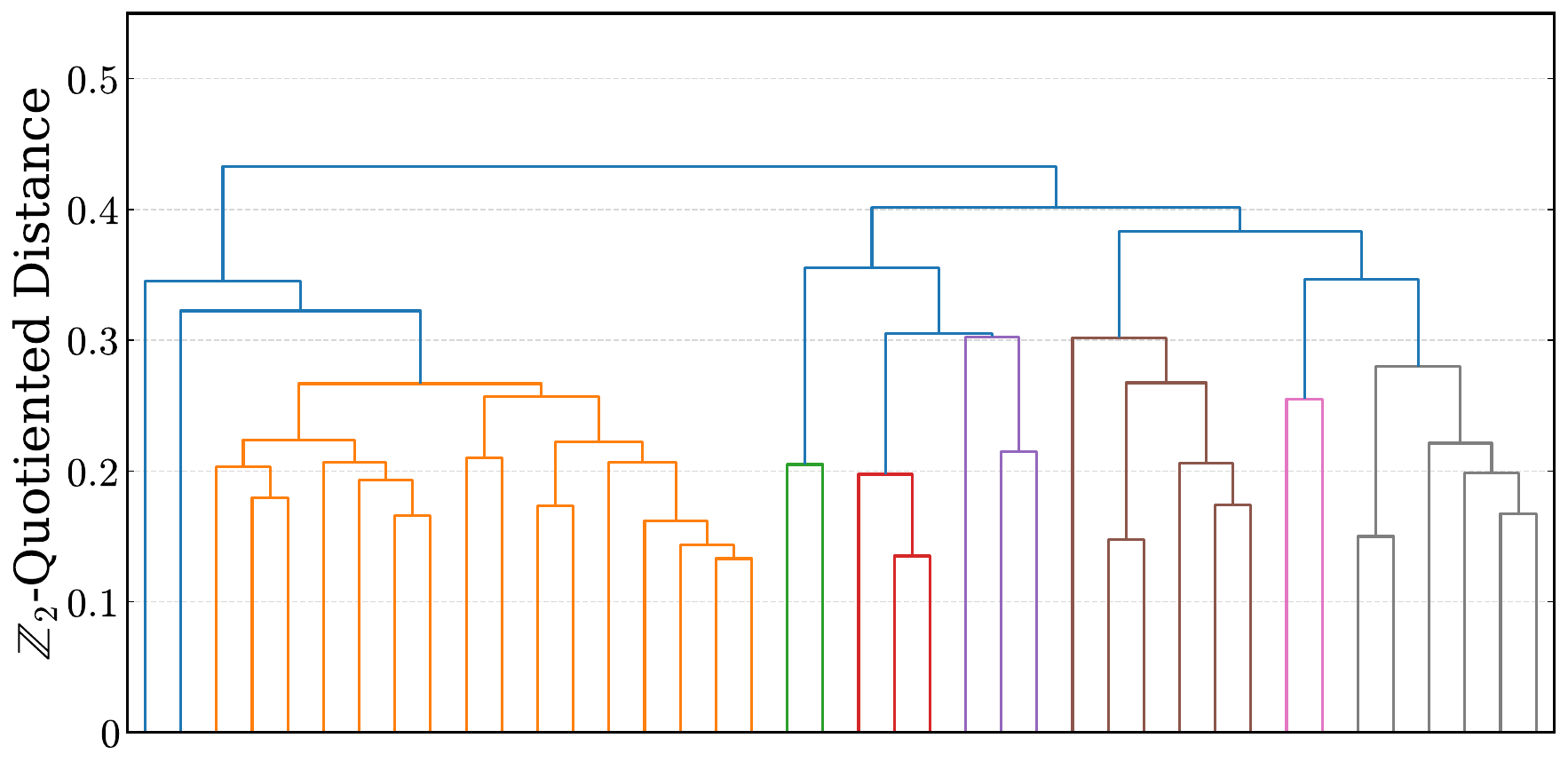}
        \caption{}
        \label{fig:traj_third}
    \end{subfigure}
    \caption{
Hierarchical clustering of the final spin configurations for the same
representative $J_0=0$ disorder realization shown in
Fig.~\ref{fig:energy_evolution}. Pairwise normalized
$\mathbb{Z}_2$-Hamming distances among all 402 replicas are clustered
using average linkage.
The dendrograms are truncated to the final 40 merged clusters for visual
clarity.
Panels show (a) the selected OIM+ population,
(b) the corresponding post-refinement DH--OIM+ population, and
(c) the independent pure-MH population.
}
    \label{fig:dendrograms}
\end{figure}

This distinction is also visible in the representative dendrograms shown in
Fig.~\ref{fig:dendrograms}. The OIM+ and DH--OIM+ ensembles exhibit more
compact within-branch clustering, with many configurations merging at smaller
$\mathbb{Z}_2$-Hamming distances, whereas the pure-MH ensemble shows broader branches and
larger characteristic merge distances. These differences indicate that the
optimization protocol influences not only the energies reached but also the
organization of the resulting configurations in Hamming-distance space.

To quantify the organization visible in the dendrograms, we consider the
cophenetic correlation coefficient together with the mean intra-cluster and
inter-cluster $\mathbb{Z}_2$-Hamming distances. As defined in Sec.~\ref{sec:model},
the cophenetic correlation measures how faithfully the hierarchical tree
represents the underlying pairwise-distance structure, while the intra- and
inter-cluster distances characterize cluster compactness and separation,
respectively.

For $J_0=0$, the cophenetic correlation coefficients are
$0.911$ for OIM+, $0.941$ for DH--OIM+, and $0.939$ for pure MH.
Thus, the pairwise-distance structure of all three ensembles is represented
well by average-linkage hierarchical clustering, with very similar
cophenetic correlations for DH--OIM+ and pure MH and a somewhat lower
value for OIM+.

For each disorder realization, pairwise normalized
$\mathbb{Z}_2$-Hamming distances are computed among all 402 configurations
and hierarchical clustering is performed using average linkage. Rather than
imposing a fixed distance threshold, we identify the largest gap between
successive linkage heights and place the clustering threshold at the midpoint
of this gap. The resulting labels define the clusters used to calculate the
mean intra- and inter-cluster distances.

For the representative realization shown in
Fig.~\ref{fig:dendrograms}, the $\mathbb{Z}_2$-quotiented hierarchy
separates into two dominant branches. Because configurations related by a
global spin inversion are identified under $d_{\mathbb{Z}_2}$, this
separation cannot be attributed to the trivial $\mathbb{Z}_2$ symmetry of
the zero-field SK Hamiltonian. Instead, it reflects configurational
structure that remains after quotienting out the global inversion symmetry.
The smaller intra-cluster distances of the OIM+ ensembles indicate more
compact groups of related configurations, while the inter-cluster
distances characterize the separation between these groups.

\begin{table*}[t]
\caption{
Disorder-averaged clustering metrics for OIM+, DH--OIM+, and pure MH
at $J_0=0$ and $J_0=-4$, averaged over 30 independent disorder
realizations. The cophenetic correlation quantifies how faithfully the
average-linkage dendrogram represents the pairwise $\mathbb{Z}_2$-Hamming-distance
structure. Inter- and intra-cluster $\mathbb{Z}_2$-Hamming distances characterize,
respectively, the separation between clusters and the compactness of
configurations within clusters.}
\label{tab:distance_metrics}
\begin{ruledtabular}
\begin{tabular}{llccc}
$J_0$ & Algorithm & Cophenetic correlation &
Inter-cluster $d_{\mathbb{Z}_2}$ & Intra-cluster $d_{\mathbb{Z}_2}$ \\
\hline
\multirow{3}{*}{$0$}
 & OIM+      & 0.911 & 0.349 & 0.142 \\
 & DH--OIM+  & 0.941 & 0.350 & 0.126 \\
 & Pure MH   & 0.939 & 0.402 & 0.154 \\
\hline
\multirow{3}{*}{$-4$}
 & OIM+      & 0.900 & 0.343 & 0.149 \\
 & DH--OIM+  & 0.930 & 0.340 & 0.134 \\
 & Pure MH   & 0.937 & 0.409 & 0.166 \\
\end{tabular}
\end{ruledtabular}
\end{table*}

For $J_0=0$, the pure-MH ensemble has a mean inter-cluster
$\mathbb{Z}_2$-Hamming distance of $0.402$ and a mean intra-cluster
distance of $0.154$. For OIM+, these values are $0.349$ and $0.142$,
while for DH--OIM+ they are $0.350$ and $0.126$, respectively.
The corresponding inter- to intra-cluster distance ratios are
approximately $2.46$ for OIM+, $2.77$ for DH--OIM+, and $2.61$ for pure
MH. Thus, DH--OIM+ exhibits the most compact within-cluster organization
and the strongest separation relative to cluster width, while pure MH has
the largest absolute inter-cluster distance. The diagnostic MH refinement
reduces the intra-cluster distance relative to OIM+ while leaving the
inter-cluster distance nearly unchanged.

These quantitative differences are also visible in the representative
dendrograms in Fig.~\ref{fig:dendrograms}. Consistent with their smaller
intra-cluster $\mathbb{Z}_2$-Hamming distances, the OIM+ and DH--OIM+
ensembles form more compact branches that merge internally at smaller
distances. This visual structure is consistent with the smaller
intra-cluster distances of the OIM-based ensembles. In particular,
DH--OIM+ exhibits the most compact within-cluster organization and the
largest inter- to intra-cluster distance ratio among the three methods
considered.

These observations suggest that the oscillator dynamics does not simply
terminate at a higher-energy version of the same ensemble. Instead, OIM+
produces a distinct organization in configuration space, and the subsequent
diagnostic MH refinement modifies this structure only modestly while further
lowering the energy. This combination suggests that the refinement may act
through relatively local changes rather than a wholesale reorganization of
the ensemble. We therefore examine which spins are altered by the diagnostic
MH sweep and whether they share a common local-stability signature.

\begin{figure*}[t]
    \centering

    \begin{subfigure}[b]{0.49\textwidth}
        \centering
        \includegraphics[width=\linewidth]
        {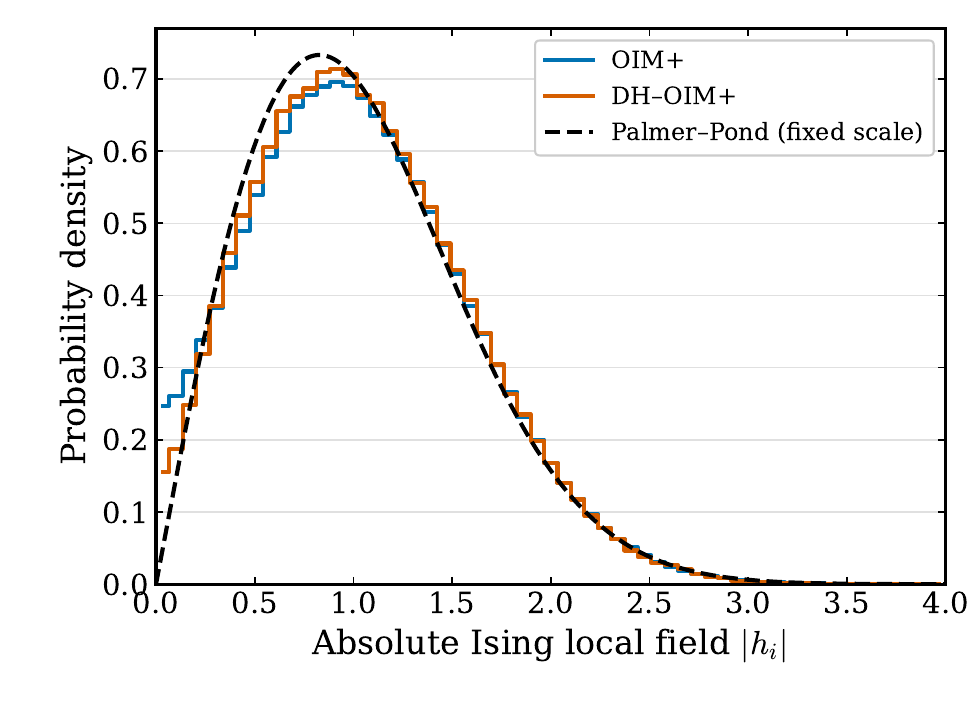}
        \caption{}
        \label{fig:field_palmer}
    \end{subfigure}
    \hfill
    \begin{subfigure}[b]{0.49\textwidth}
        \centering
        \includegraphics[width=\linewidth]
        {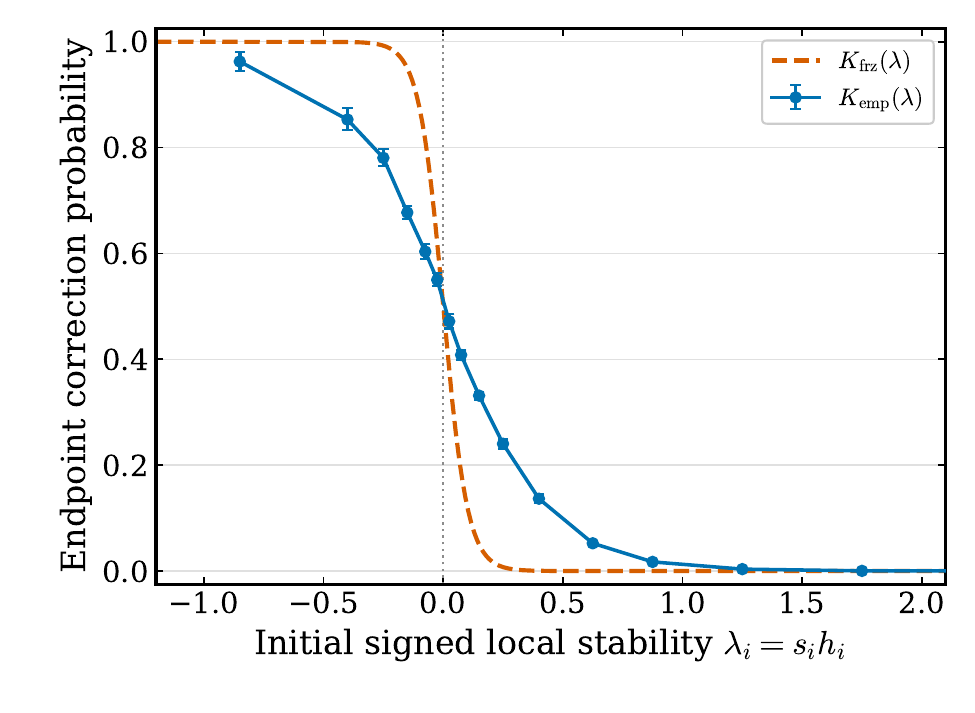}
        \caption{}
        \label{fig:field_kernel}
    \end{subfigure}

\caption{
Local-field structure and defect correction for the $J_0=0$ SK
ensemble, pooled over 30 independent disorder realizations with
402 replicas per realization.
(a) Distribution of the absolute Ising local-field magnitude $|h_i|$
for the selected OIM+ population and the corresponding DH--OIM+
population after diagnostic MH refinement. The fixed Palmer--Pond
reference is rescaled to the coupling normalization used here,
$\sigma_{\rm PP}=1.17/\sqrt{2}$. Both populations exhibit a broad
distribution close to this finite-size SK reference, while the
refinement robustly suppresses the excess weight at small $|h_i|$.
(b) Probability that a spin ends the refinement with the opposite
orientation as a function of its pre-refinement signed local stability,
$\lambda_i=s_i h_i$. The measured correction probability
$K_{\rm emp}(\lambda)$ decreases strongly with increasing $\lambda$ and
is compared with the frozen-field prediction $K_{\rm frz}(\lambda)$,
obtained by applying the same time-dependent MH schedule while holding
the initial local field fixed. The error bars on $K_{\rm emp}(\lambda)$ denote 95\% bootstrap
confidence intervals obtained by resampling the 30 disorder realizations. The frozen-field model captures the
qualitative dependence on local stability but does not reproduce the
measured correction probabilities quantitatively.
}
\label{fig:field_distribution}
\end{figure*}

\begin{figure}[t!]
    \centering

    \begin{subfigure}[b]{\linewidth}
        \centering
        \includegraphics[width=\textwidth]{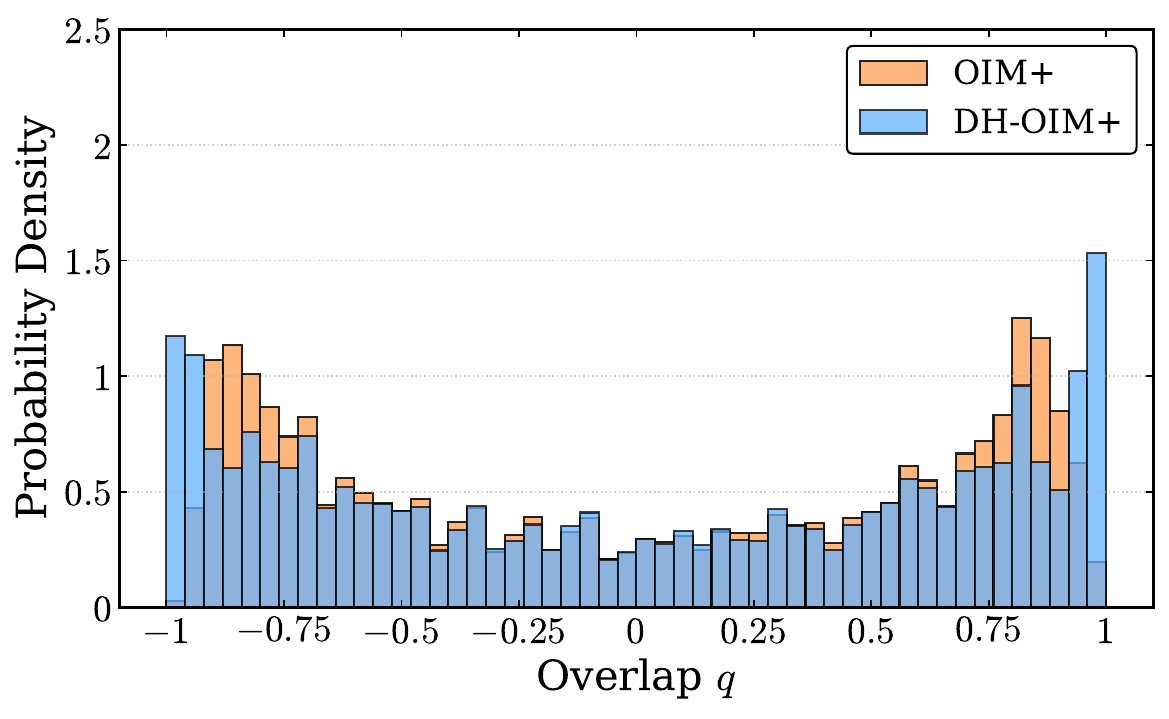}
        \caption{}
        \label{fig:pq_J0m4_oim}
    \end{subfigure}

    \vspace{0.1cm}

    \begin{subfigure}[b]{\linewidth}
        \centering
        \includegraphics[width=\textwidth]{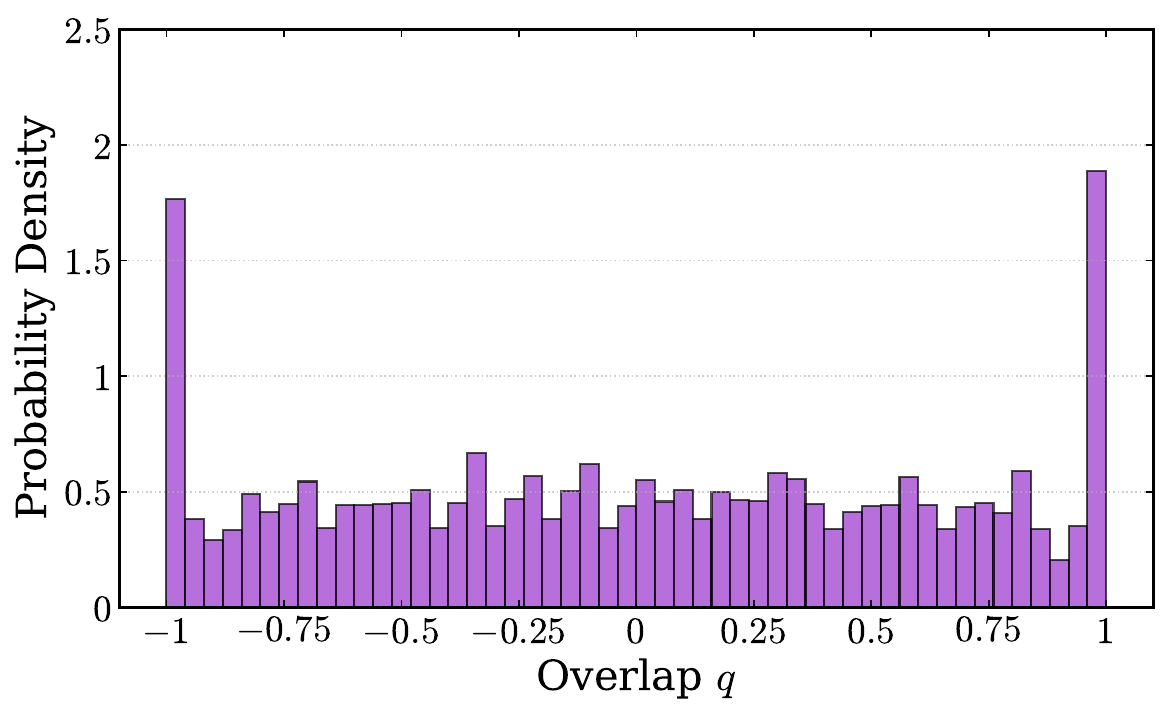}
        \caption{}
        \label{fig:pq_J0m4_mh}
    \end{subfigure}

    \caption{
    Disorder-averaged overlap distributions $P(q)$ for the
    antiferromagnetically biased SK ensemble with $J_0=-4$, averaged over
    30 independent disorder realizations. For each realization, overlaps
    are computed between replica configurations generated for the same
    coupling matrix $J$, and the resulting distributions are subsequently
    averaged over disorder.
    (a) Comparison of the selected OIM+ population and the corresponding
    post-refinement DH--OIM+ population.
    (b) Overlap distribution for the independent pure-MH baseline.
    All three ensembles retain substantial probability density over
    intermediate overlap values, while differences in the relative
    concentration near $q=\pm1$ indicate protocol-dependent organization
    of the resulting configurations.
    }
    \label{fig:pq_J0m4}
\end{figure}

\begin{figure}[t!]
    \centering

    \begin{subfigure}[b]{\linewidth}
        \centering
        \includegraphics[width=\textwidth]{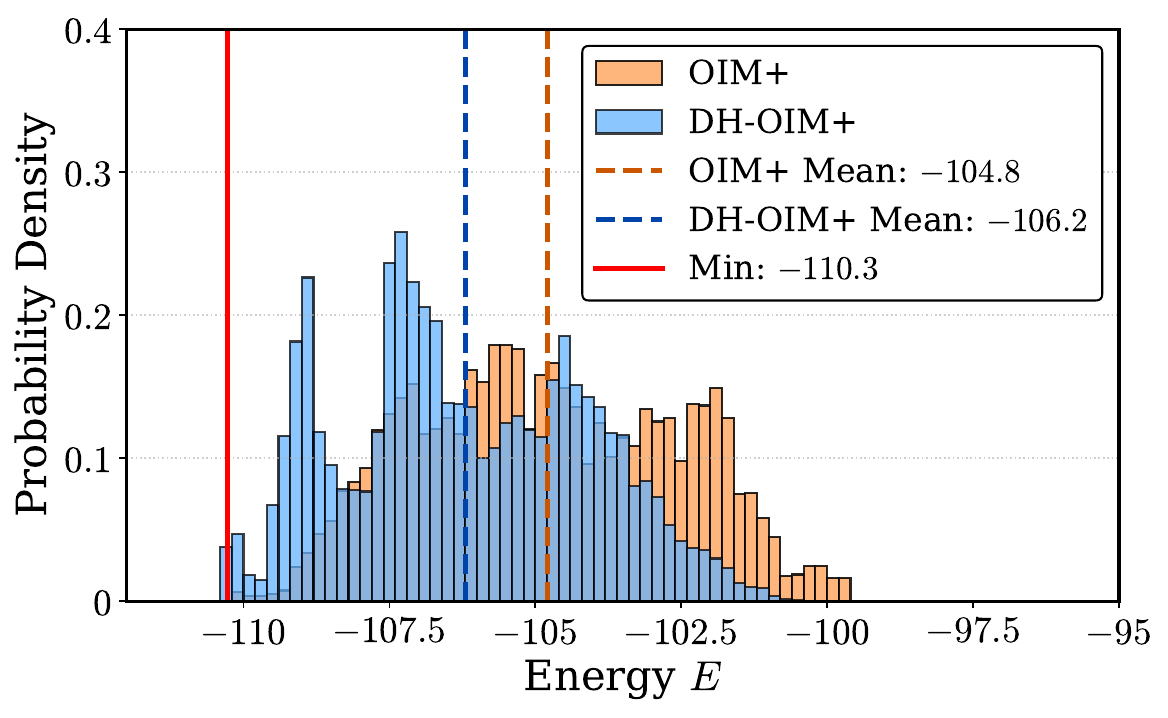}
        \caption{}
        \label{fig:afm_energy_oim}
    \end{subfigure}

    \vspace{0.1cm}

    \begin{subfigure}[b]{\linewidth}
        \centering
        \includegraphics[width=\textwidth]{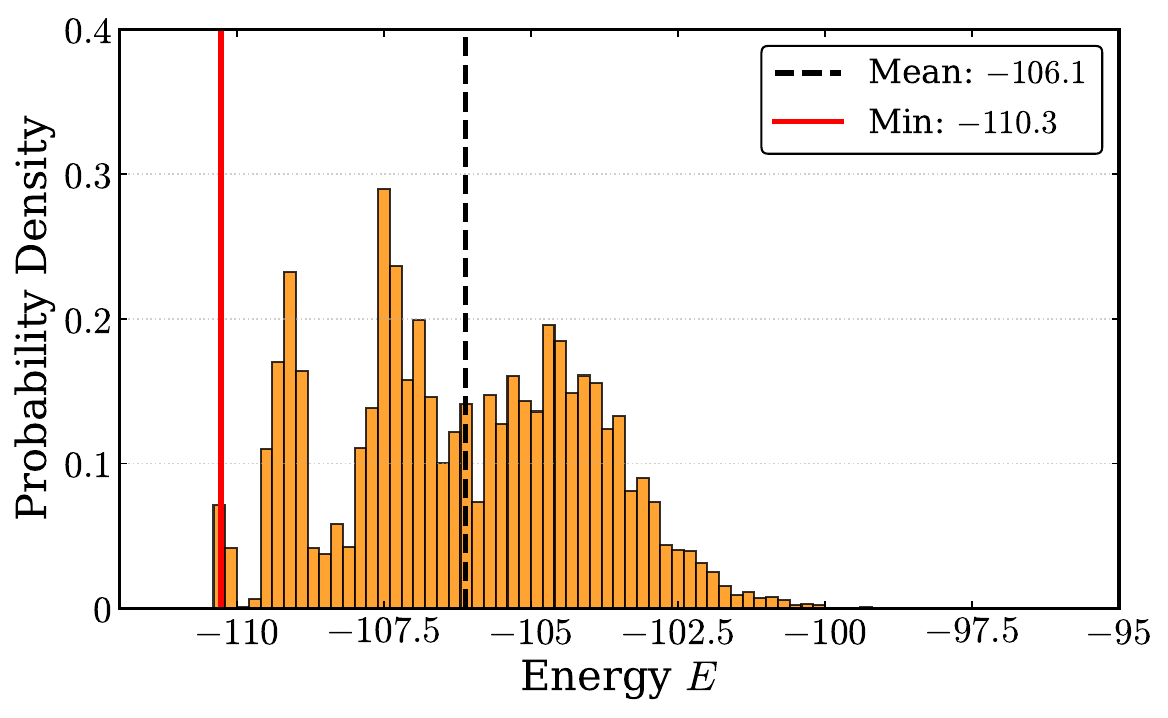}
        \caption{}
        \label{fig:afm_energy_mh}
    \end{subfigure}

    \caption{
    Distribution of final Ising energies for the antiferromagnetically
    biased SK ensemble with $J_0=-4$, pooled over 30 independent disorder
    realizations with 402 replicas per realization ($12\,060$
    configurations for each distribution).
    (a) Final energies of the selected OIM+ population before and after
    the short diagnostic MH refinement. The refinement shifts the pooled
    mean energy from $-104.8$ for OIM+ to $-106.2$ for DH--OIM+ and
    suppresses the higher-energy tail.
    (b) Final-energy distribution obtained from the independent pure-MH
    baseline over the same disorder ensemble, with pooled mean energy
    $-106.1$. The lowest observed energy is $-110.3$ for both
    DH--OIM+ and pure MH.
    }
    \label{fig:afm_energies}
\end{figure}

\subsection{Weak-field residual defects and local-stability models}
\label{sec:weak_field}

The preceding analysis shows that the diagnostic MH refinement lowers the
energy of the selected OIM+ population while producing only modest changes
in its broader configuration-space organization. We therefore turn to the
local Ising stability of the post-binarization configurations in order to
identify which residual degrees of freedom remain susceptible to local
correction.

For the noiseless oscillator dynamics at fixed $K_s$, the Lyapunov
function in Eq.~\ref{eq:lyapunov} is nonincreasing along the deterministic
trajectory. This gradient-like relaxation does not, however, guarantee
convergence to the global minimum, and the annealing dynamics can terminate
in configurations whose binarized spin states remain locally improvable.

For a binarized configuration $\mathbf{s}$, we define the local field
\begin{equation}
    h_i = \sum_{j\neq i} J_{ij}s_j ,
    \label{eq:local_field}
\end{equation}
so that the Ising energy may be written as
\begin{equation}
    E(\mathbf{s})
    =
    -\frac{1}{2}\sum_i s_i h_i .
    \label{eq:energy_local_field}
\end{equation}
A reversal of spin $i$ changes the energy by
\begin{equation}
    \Delta E_i = 2s_i h_i .
    \label{eq:single_flip_energy}
\end{equation}
It is therefore useful to introduce the signed local stability
\begin{equation}
    \lambda_i = s_i h_i .
    \label{eq:local_stability}
\end{equation}
Sites with $\lambda_i<0$ are locally unstable to a zero-temperature
single-spin flip, while sites with small positive $\lambda_i$ are locally
stable but only weakly so. At finite $T_{\rm MH}$, energetically favorable
flips with $\lambda_i<0$ are accepted, while flips with small positive
$\lambda_i$ can also occur with appreciable probability. Thus, the
diagnostic refinement is expected to act preferentially on sites of small
signed local stability. The magnitude $|h_i|=|\lambda_i|$ measures the
strength of the local energetic bias independently of the current spin
orientation.

The depletion of very small local fields is a characteristic property of
low-energy SK states. Palmer and Pond showed that stability of the Ising
ground state requires the small-field density to vanish at least linearly,
$P(h)=O(h)$, and their finite-$N$ simulations displayed an approximately
linear rise from a nonzero finite-size intercept \cite{Palmer1979}.
For $N=200$ they further found that the broad local-field distribution is
reasonably described by the empirical form
\begin{equation}
    P_{\rm PP}(h)
    =
    \frac{h}{\sigma_{\rm PP}^{\,2}}
    \exp\!\left(
        -\frac{h^2}{2\sigma_{\rm PP}^{\,2}}
    \right),
    \qquad h\geq0 ,
    \label{eq:palmer_distribution}
\end{equation}
with $\sigma_{\rm PP}=1.17$ in their coupling normalization. This expression
should be regarded as a finite-size empirical reference rather than an
exact form of the thermodynamic-limit SK distribution.

The coupling variance used by Palmer and Pond is $1/N$, whereas the
symmetrized couplings used here have variance $1/(2N)$. Local fields
therefore differ by a factor $1/\sqrt{2}$, giving the parameter-free
reference scale
\begin{equation}
    \sigma_{\rm PP}
    =
    \frac{1.17}{\sqrt{2}}
    \approx
    0.827 .
    \label{eq:palmer_rescaling}
\end{equation}

Figure~\ref{fig:field_distribution}(a) compares this fixed reference with
the measured $P(|h|)$ for OIM+ and DH--OIM+. The broad OIM+ distribution is
already close to the finite-size SK reference. An unbinned fit of
Eq.~\eqref{eq:palmer_distribution} gives
$\hat{\sigma}=0.832$, approximately $0.6\%$ above the rescaled Palmer value.
The corresponding empirical mean field,
$\langle |h|\rangle=1.037$, is likewise nearly identical to the fixed
Palmer prediction, $1.037$.

The principal discrepancy occurs near the origin. For the OIM+ population,
a linear fit over $0\leq |h|\leq0.4$ gives
\begin{equation}
    P_{\rm OIM+}(h)
    \simeq
    0.211 + 0.569 h ,
    \label{eq:oim_small_field}
\end{equation}
revealing appreciable finite-$N$ weight at weak local fields. The
diagnostic MH refinement robustly suppresses this excess. Over the same
range the fitted intercept decreases to approximately $0.105$.
The decrease in the low-field intercept is reproduced for all tested fitting ranges and histogram resolutions.

The reduction of weak-field weight is also visible directly in the
cumulative distributions, with the fraction of sites satisfying
$|h_i|<0.1$ decreasing from approximately $2.50\%$ in OIM+ to
$1.60\%$ after refinement.

To quantify directly which sites are susceptible to the diagnostic MH
refinement, we measure the probability that a spin ends the refinement with
the opposite orientation as a function of its initial signed local
stability,
\begin{equation}
    \lambda_i=s_i h_i .
\end{equation}
We denote this empirical correction probability by
\begin{equation}
    K_{\rm emp}(\lambda)
    =
    P(\text{spin changed after MH}\mid\lambda).
    \label{eq:correction_kernel}
\end{equation}
Figure~\ref{fig:field_distribution}(b) shows that
$K_{\rm emp}(\lambda)$ decreases strongly with increasing $\lambda$.
Spins that are initially unstable, $\lambda<0$, are therefore especially
likely to be corrected, while weakly stable spins with small positive
$\lambda$ also retain an appreciable probability of changing.

Across the full $J_0=0$ ensemble, the refinement changes
$142700$ of $2412000$ spin endpoints, corresponding to approximately
$5.92\%$ of all sites. Among initially unstable spins,
$65.8\%$ are changed by the refinement, compared with only $4.74\%$ of
initially stable spins. Initially unstable sites constitute only $1.93\%$
of the OIM+ population, however. Consequently, although an unstable site
is far more likely to be corrected, most corrected endpoints
($78.6\%$) originate from the much larger population of initially stable
spins.

Thus, initially unstable spins are much more likely to be corrected, but
they account for only a minority of all corrected endpoints. The refinement
also modifies initially stable spins, particularly those with small positive
local stability.
The strong dependence of the correction probability on the initial
stability raises a simple question: can
$K_{\rm emp}(\lambda)$ be explained solely by the Metropolis acceptance
rule applied to a spin in its initial local field?

To test this, we construct a frozen-field reference model in which the
neighbors of a given spin are held fixed throughout the refinement.
Its signed local stability $\lambda=s_i h_i$ therefore remains constant.
If the spin is in its initial orientation, reversing it changes the energy
by $2\lambda$, while reversing it back changes the energy by $-2\lambda$.
The corresponding Metropolis acceptance probabilities are
\begin{align}
a_{+}(\lambda,T)
&=
\min\!\left[1,e^{-2\lambda/T}\right],\\
a_{-}(\lambda,T)
&=
\min\!\left[1,e^{2\lambda/T}\right].
\end{align}

Let $q(u;\lambda)$ denote the probability that the spin is opposite to its
initial orientation after $u$ proposal opportunities. In the frozen-field
model,
\begin{equation}
\frac{dq}{du}
=
a_{+}(\lambda,T(u))[1-q]
-
a_{-}(\lambda,T(u))q,
\quad q(0)=0 .
\label{eq:frozen_field_ode}
\end{equation}
The diagnostic refinement contains $10^4$ proposals for $N=200$ spins, so
each spin is selected 50 times on average. Expressed in this proposal clock,
the temperature decreases from $0.2$ to $0.1$ according to
\begin{equation}
T(u)=0.2-\frac{0.1}{50}u,
\qquad 0\leq u\leq50 .
\end{equation}
The predicted probability that the spin ends the refinement reversed is
therefore
\begin{equation}
K_{\rm frz}(\lambda)=q(50;\lambda).
\end{equation}

If the initial local field were sufficient to determine the subsequent
correction, $K_{\rm frz}(\lambda)$ should reproduce the measured
$K_{\rm emp}(\lambda)$. It reproduces the qualitative trend---low-stability
spins are more likely to change---but not the measured probabilities.
In particular, the frozen-field model predicts too few endpoint changes
among initially stable spins and too many among initially unstable spins.

This discrepancy has a natural origin. In the actual MH refinement, the
neighbors are not frozen: whenever another spin flips, the local field
$h_i=\sum_jJ_{ij}s_j$ changes. A spin that is initially stable can therefore
become weakly stable or unstable later in the sweep, while a spin that is
initially unstable can flip and subsequently return to its original
orientation. Thus, the initial stability identifies where the refinement is
most likely to act, but it is not sufficient to determine the final spin
state.

Finally, a small net local field may arise either because the couplings
incident on a site are collectively weak or because substantial competing
interaction terms nearly cancel. To distinguish these possibilities, we
define
\begin{equation}
    L_i
    =
    \sum_{j\neq i}|J_{ij}|,
    \label{eq:incident_strength}
\end{equation}
and
\begin{equation}
    r_i
    =
    \frac{|h_i|}{L_i}.
    \label{eq:cancellation_ratio}
\end{equation}
Here $L_i$ measures the total incident coupling magnitude, whereas $r_i$
measures the residual field relative to that interaction scale. A small
$|h_i|$ accompanied by an ordinary $L_i$ and a small $r_i$ therefore
signals cancellation among competing interaction contributions rather
than an intrinsically weak coupling environment.

The disorder-averaged values are summarized in
Table~\ref{tab:weak_field_metrics}. For $J_0=0$, corrected spins have
$\langle |h_i|\rangle=0.32\pm0.03$, compared with
$1.08\pm0.02$ for unchanged spins. Their total incident coupling
magnitudes, however, are nearly the same:
$L_i=7.91\pm0.06$ and $7.96\pm0.04$, respectively.
Correspondingly, the cancellation ratio is much smaller for corrected
sites,
$r_i=0.040\pm0.004$, than for unchanged sites,
$r_i=0.136\pm0.002$.
The same ordering persists for $J_0=-4$.

\begin{table}[t]
\caption{
Disorder-averaged absolute local field $|h_i|$, total incident coupling
magnitude $L_i$, and cancellation ratio $r_i$ for spins corrected by the
diagnostic MH refinement and for unchanged spins. Values are means
$\pm$ standard deviations over 30 independent disorder realizations.
}
\label{tab:weak_field_metrics}
\centering
\begin{ruledtabular}
\begin{tabular}{c l c c}
$J_0$ & Metric & Corrected spins & Unchanged spins \\
\hline
$0$
& $|h_i|$ & $0.32\pm0.03$ & $1.08\pm0.02$ \\
& $L_i$    & $7.91\pm0.06$ & $7.96\pm0.04$ \\
& $r_i$    & $0.040\pm0.004$ & $0.136\pm0.002$ \\
\hline
$-4$
& $|h_i|$ & $0.32\pm0.03$ & $1.10\pm0.02$ \\
& $L_i$    & $8.55\pm0.09$ & $8.58\pm0.04$ \\
& $r_i$    & $0.038\pm0.004$ & $0.128\pm0.002$ \\
\end{tabular}
\end{ruledtabular}
\end{table}

Taken together, these results give a more specific interpretation of the
residual defects left by OIM+. The selective oscillator dynamics already
produces the broad local-field structure characteristic of low-energy SK
configurations, but retains an enhanced finite-size population of weakly
stabilized sites.
The short MH stage therefore acts preferentially on residual weakly
stabilized sites rather than performing a wholesale reorganization of
configuration space, consistent with the modest changes observed in the
overlap and clustering diagnostics. At the same time, the failure of the
frozen-field null model shows that this local correction is not a collection
of independent single-spin events: the surrounding local fields evolve as
the configuration is updated.

\subsection{Antiferromagnetic frustration} \label{sec:antiferro}

For $J_0=-4$, differences in the organization of the three ensembles
persist. As summarized in Table~\ref{tab:distance_metrics}, the
cophenetic correlation coefficients are $0.900$ for OIM+, $0.930$ for
DH--OIM+, and $0.937$ for pure MH, indicating that all three
$\mathbb{Z}_2$-Hamming-distance structures are represented
well by the average-linkage hierarchy, with the highest correlation for
pure MH. Both OIM+ and DH--OIM+ exhibit smaller intra-cluster distances
than pure MH, indicating more compact within-cluster organization,
whereas pure MH has the largest absolute inter-cluster distance. The
corresponding inter- to intra-cluster distance ratios are approximately
$2.29$, $2.55$, and $2.46$ for OIM+, DH--OIM+, and pure MH,
respectively, so DH--OIM+ exhibits the strongest separation relative to
cluster width.

The disorder-averaged overlap distributions for the antiferromagnetically
biased ensemble are shown in Fig.~\ref{fig:pq_J0m4}. As in the $J_0=0$
case, all three protocols retain substantial weight at intermediate
overlap values, indicating populations of distinct and partially
correlated configurations rather than collapse exclusively near
$q=\pm1$. The OIM+ and DH--OIM+ distributions remain qualitatively
similar, indicating that the short diagnostic MH refinement modifies the
overlap structure only modestly under the antiferromagnetic bias.

The same qualitative refinement behavior is observed in the final
energies. For $J_0=-4$, the pooled mean energy decreases from
approximately $-104.8$ for OIM+ to $-106.2$ after the diagnostic MH
refinement, while the independent pure-MH baseline reaches a comparable
mean energy of approximately $-106.1$. Thus, DH--OIM+ and pure MH reach comparable pooled mean energies for the
biased ensemble, while both methods attain the same observed minimum energy
of approximately $-110.3$.

\begin{table*}[t]
\caption{
Max-Cut performance of DH--OIM+ and pure MH on representative G-set
benchmark instances. In this application-oriented benchmark, the objective
is the best cut obtained by the complete optimization protocol rather than
the intermediate ensemble structure, so the refined DH--OIM+ output is used
as the final OIM-based solution. ``Best known'' denotes the reference cut
value for each instance compiled by Goudet
\emph{et al.} \cite{Goudet2024}, and relative performance is the achieved cut value
divided by the best-known value.
}
\label{tab:gset_results}
\begin{ruledtabular}
\begin{tabular}{lccccccccc}
Instance &
Nodes &
Edges &
Weights &
Class &
Best Known&
DH--OIM+&
MH&
DH--OIM+/ best known&
MH/ best known
\\
\hline
G1  & 800  & 19176 & $+1$     & Random   & 11624 & 11624 & 11550 & 100.00\% & 99.36\% \\
G6  & 800  & 19176 & $\pm1$   & Random   & 2178  & 2177  & 2113  & 99.95\% & 97.02\% \\
G11 & 800  & 1600  & $\pm1$   & Toroidal & 564   & 562   & 562   & 99.65\% & 99.65\% \\
G14 & 800  & 4694  & $+1$     & Planar   & 3064  & 3062  & 3040  & 99.93\% & 99.22\% \\
G15 & 800  & 4661  & $+1$     & Planar   & 3050  & 3050  & 3023  & 100.00\% & 99.11\% \\
G22 & 2000 & 19990 & $+1$     & Random   & 13359 & 13357 & 13161 & 99.99\% & 98.52\% \\
G23 & 2000 & 19990 & $+1$     & Random   & 13344 & 13339 & 13160 & 99.96\% & 98.62\% \\
G36 & 2000 & 11766 & $+1$     & Planar   & 7680  & 7672  & 7572 & 99.90\% & 98.59\% \\
G39 & 2000 & 11778 & $\pm1$   & Planar   & 2408  & 2407  & 2266 & 99.96\% & 94.10\% \\
\end{tabular}
\end{ruledtabular}
\end{table*}

Figure~\ref{fig:afm_energies} compares the final-energy distributions of
OIM+, DH--OIM+, and pure Metropolis--Hastings annealing for $J_0=-4$.
The qualitative behavior observed for $J_0=0$ persists under the
antiferromagnetic bias. OIM+ already reaches low-energy configurations,
while the subsequent diagnostic MH refinement shifts the ensemble further
toward lower energies, reducing the pooled mean from approximately $-104.8$
to $-106.2$. The independent pure-MH baseline reaches a comparable mean
energy of approximately $-106.1$, and both DH--OIM+ and pure MH attain the
same lowest observed energy of approximately $-110.3$.

The persistence of the same qualitative refinement behavior under
$J_0=-4$ is consistent with the weak-field defect picture developed above.
Despite the antiferromagnetic bias, OIM+ again produces a low-energy
population that is further improved by the short diagnostic MH sweep,
while the final DH--OIM+ and pure-MH energies remain comparable.
The local-field analysis in Table~\ref{tab:weak_field_metrics} shows that
the same pattern also persists at $J_0=-4$: corrected spins have
substantially smaller $|h_i|$ and $r_i$ than unchanged spins, while their
incident coupling magnitudes $L_i$ remain comparable. Thus, the residual
weak fields under antiferromagnetic bias are again associated primarily
with cancellation among competing interactions rather than unusually weak
coupling environments.

Taken together, these results show that the physical picture developed for
the unbiased SK ensemble is not specific to $J_0=0$. Under the
antiferromagnetic bias, OIM+ again generates a structured low-energy
population, while the short diagnostic MH refinement primarily corrects
residual weak-field spins and leaves the broader configuration-space
organization largely intact. The resulting DH--OIM+ energies are comparable
to those of the independent pure-MH baseline.  Thus, the separation between collective oscillator exploration and
subsequent local defect correction persists despite the substantial
antiferromagnetic frustration introduced by the biased coupling
distribution.

\section{G-set benchmark graphs}
\label{sec:benchmarks}

The preceding SK analysis was designed primarily to resolve the physical
behavior of the selective oscillator dynamics by examining OIM+ separately
from the subsequent diagnostic MH refinement. The G-set benchmark serves a
different purpose: here we assess the application-level optimization
performance of the complete protocol. We therefore treat DH--OIM+ as the
final OIM-based solver and compare its achieved cut values directly with
those of the independent pure-MH baseline. We evaluate DH--OIM+ on selected
instances from the standard G-set Max-Cut benchmark suite
\cite{Helmberg2000}, using the best-known cut values compiled by Goudet
\emph{et al.} \cite{Goudet2024} as the reference.

To test whether this end-to-end protocol remains competitive beyond the
fully connected SK ensemble, we consider nine representative G-set Max-Cut
instances spanning different graph sizes, weight structures, and graph
classes, with system sizes ranging from $N=800$ to $N=2000$.

In the Max-Cut problem, the vertices of a weighted graph are partitioned
into two sets so as to maximize the total weight of the edges connecting
the two sets. Assigning an Ising spin $s_i=\pm1$ to each vertex, the cut
value can be written as~\cite{Lucas_2014,OIM}
\begin{equation}
C(\mathbf{s})
=
\sum_{(i,j)\in E}
W_{ij}\frac{1-s_i s_j}{2}
=
\frac{1}{2}W_{\mathrm{total}}
-
\frac{1}{2}
\sum_{(i,j)\in E}W_{ij}s_i s_j ,
\end{equation}
where
\begin{equation}
W_{\mathrm{total}}
=
\sum_{(i,j)\in E}W_{ij}.
\end{equation}
With the identification $J_{ij}=-W_{ij}$, the corresponding Ising energy is
\begin{equation}
E(\mathbf{s})
=
-\sum_{(i,j)\in E}J_{ij}s_i s_j
=
\sum_{(i,j)\in E}W_{ij}s_i s_j ,
\end{equation}
and hence
\begin{equation}
C(\mathbf{s})
=
\frac{1}{2}W_{\mathrm{total}}
-
\frac{1}{2}E(\mathbf{s}).
\end{equation}
Thus, maximizing the Max-Cut objective is equivalent to minimizing the
corresponding Ising energy.

For the G-set calculations, the system size and coupling matrix are determined
by the corresponding benchmark instance. Both DH--OIM+ and the independent
pure-MH baseline use 448 replicas per instance, providing a population-matched
comparison. For the post-OIM+ MH refinement, each replica undergoes
$10^4$ single-spin Metropolis proposals while the temperature is annealed
linearly from $T_{\rm MH}=0.1$ to $0.01$. The pure-MH benchmark follows the
annealing, settling, and measurement protocol described in Sec.~\ref{sec:healing}, with
the final temperature likewise set to $T_{\rm MH}=0.01$.

Table~\ref{tab:gset_results} summarizes the resulting Max-Cut performance.
Across the nine tested instances, DH--OIM+ attains cut values close to the
best-known solutions and equals or exceeds the pure-MH result in every
case. In particular, DH--OIM+ reaches the best-known cut for G15 and
achieves more than $99\%$ of the best-known value for all nine instances.
The evolution of the G-set populations across the four oscillator epochs
is summarized in Appendix~\ref{app:Intermediate}.

The digital refinement following OIM+ uses only $10^4$ single-spin
Metropolis proposals per replica, compared with $1.2\times10^5$ proposals
per replica for the independent pure-MH baseline. Thus, within this
end-to-end comparison, high-quality OIM-based solutions are obtained with
only a comparatively short subsequent sequence of local Metropolis updates.
This reduced reliance on sequential digital refinement is particularly
relevant to the hybrid analog--digital implementation discussed in
Sec.~\ref{sec:disc}.

The G-set benchmarks therefore show that the end-to-end performance of
DH--OIM+ is not restricted to the fully connected SK setting. At the same
time, these benchmark results do not establish that the weak-field defect
mechanism identified in the SK ensembles is responsible for the observed
performance on G-set graphs. Rather, they demonstrate that selective
oscillator evolution followed by a short local refinement remains effective
across a more heterogeneous set of Ising optimization problems.

\section{Discussion}
\label{sec:disc}

The SK analysis reveals a clear separation between collective exploration and residual local correction in the selective oscillator dynamics. Taken together, the energy, overlap, distance, clustering, and local-field diagnostics show that OIM+ organizes a nontrivial ensemble of distinct low-energy configurations before any local digital refinement is applied, complementing recent observations that physical annealers can access structured collections of deep low-energy states \cite{Zhang2024Cyclic}. The remaining imperfections are comparatively local: the subsequent diagnostic MH sweep modifies only a small fraction of spins, concentrated predominantly at weak post-binarization local stabilities.

An additional consequence is that strong replica selection need not reduce
the search to repeated copies of a single low-energy state. Despite the
repeated elimination and repopulation steps, the final OIM+ population
retains substantial intermediate-overlap weight and nontrivial
configuration-space organization. Selection therefore improves the
energetic composition of the population without, in the present
calculations, eliminating its broader configurational diversity, preserving
access to alternative high-quality solutions that may differ in structural
properties or compatibility with additional constraints.

The local-field analysis provides a microscopic basis for this separation
of roles. The spins most susceptible to correction are those with small
local stability, and these weak fields arise predominantly from
cancellation among competing interaction terms rather than from uniformly
weak couplings. The residual limitation of the oscillator dynamics is
therefore associated with locally balanced, frustration-induced degrees of
freedom. At the same time, the comparatively small fraction of corrected
spins and the limited change in the broader overlap and
$\mathbb{Z}_2$-Hamming-distance structure indicate that the diagnostic MH
stage acts primarily as a local correction rather than restarting the
global search.

In this sense, the ensemble-level and local diagnostics play complementary
roles. The former characterize what the collective oscillator search has
already organized, whereas the latter identify the residual degrees of
freedom that it has not fully resolved.

This division of labor is particularly promising for a hybrid
analog--digital implementation. A physical OIM evolves all oscillator
phases continuously and in parallel, allowing the collective search to be
performed by the intrinsic dynamics of the hardware rather than through a
long sequence of explicit spin updates. Demonstrated OIM implementations
already operate on very short physical time scales: a 1968-node CMOS
ring-oscillator system operating near $1\,\mathrm{GHz}$ reached its reported
ground-state configuration within fewer than 50 oscillator cycles
($<50\,\mathrm{ns}$), while MEMS and integrated optoelectronic
implementations exhibit characteristic synchronization or spin-evolution
times of approximately $90$--$150\,\mathrm{ns}$
\cite{Energy4,Deng2024,Wu2025}. These figures illustrate the potential
speed of the collective analog stage, although they should not be identified
directly with the numerical integration time used here.

In a hardware implementation following this protocol, the digital intervention associated with selection and repopulation would occur only at the boundaries between annealing epochs, where the oscillator states are read out and binarized, their Ising energies evaluated, the higher-energy replicas discarded, and the surviving states used to reinitialize the next epoch. For comparison, population annealing uses temperature-dependent resampling to maintain an approximately equilibrium population during annealing~\cite{Wang2015PopulationAnnealing}. OIM+ instead uses a simpler energy-ranked culling and replication step, without requiring the replica population to represent an equilibrium distribution. This makes the digital intervention particularly straightforward to interleave with the analog oscillator dynamics. The final defect-healing stage likewise requires only a comparatively short sequence of sequential Metropolis updates. The resulting architecture therefore delegates the large-scale exploration to fast, parallel oscillator dynamics while reserving digital computation for intermittent selection, state reinitialization, and targeted local polishing. On the tested G-set instances, this division of labor retains high optimization quality while requiring a substantially shorter sequential digital refinement stage than the independent pure-MH baseline.

The use of the fully connected SK model in the mechanistic analysis should
be interpreted as a controlled testbed rather than as a requirement that
practical oscillator hardware realize dense connectivity in every
application. Fully connected interactions have been demonstrated in both
coherent and coupled-oscillator Ising-machine architectures
\cite{machines2,Lo2023}, while the G-set results show that the complete
protocol also remains effective on more heterogeneous graph structures.

Whether this architectural promise translates into a net wall-clock-time or energy advantage will depend on the overheads associated with readout, energy evaluation, replica selection, state reinitialization, and control, and ultimately requires direct hardware benchmarking. We also note that the present configuration-space clustering should be interpreted geometrically: it characterizes the organization of the sampled ensembles, in a manner related to recent basin- and cluster-based analyses of low-energy annealer outputs \cite{Zhang2025Complexity}, but does not by itself establish a correspondence with distinct dynamical basins or thermodynamic pure states.

\section{Conclusions}
\label{sec:conc}

We used ensemble-level and local-stability diagnostics to determine what
selective oscillator dynamics accomplishes in a frustrated Ising landscape
and what remains unresolved after the collective search. OIM+ drives the
replica population into low-energy regions while retaining a nontrivial
organization of distinct configurations despite repeated energy-based
selection and repopulation.

The residual corrections exposed by a short diagnostic
Metropolis--Hastings refinement are strongly concentrated at weak
post-binarization local fields. These weak fields arise predominantly from
cancellation among competing interaction terms rather than from uniformly
weak couplings. The refinement therefore lowers the energy by modifying
only a small fraction of spins while leaving the broader overlap and
$\mathbb{Z}_2$-Hamming-distance organization largely intact.

These observations support a separation of roles: the selective oscillator
dynamics performs the collective exploration and organization of the
low-energy population, whereas the subsequent local-update stage corrects
residual weak-field degrees of freedom. The same qualitative picture
persists for the antiferromagnetically biased SK ensemble with $J_0=-4$.
The G-set calculations address a complementary question and show that the
complete DH--OIM+ protocol retains competitive end-to-end optimization
performance beyond the fully connected SK setting.

The broader implication is that the value of oscillator Ising machines
need not be judged solely by the minimum energy reached in a single run.
Their collective dynamics can instead be understood in terms of the
low-energy ensembles they generate and the character of the residual
degrees of freedom they leave unresolved. This perspective motivates
hybrid architectures in which collective analog exploration is combined
with intermittent selection and targeted digital refinement.

\begin{table*}[t]
	\caption{
		Intermediate Max-Cut performance of representative G-set instances across
		the four oscillator epochs and after the subsequent defect-healing (DH)
		refinement. For each epoch, we report the maximum cut, the population mean,
		and the cutoff for the best-performing third of the 448 replicas. ``Best''
		denotes the best-known reference value compiled by Goudet \emph{et al.}~	 \cite{Goudet2024}.
	}
	\label{tab:gset_epoch_cuts}
	\centering
	\footnotesize
	\setlength{\tabcolsep}{3.2pt}
	\resizebox{\textwidth}{!}{%
		\begin{ruledtabular}
			\begin{tabular}{c c r r r r @{\hspace{8mm}} c c r r r r}

				Graph & Stage & Max-Cut & Mean & Top-third threshold & Best &
				Graph & Stage & Max-Cut & Mean & Top-third threshold & Best \\
				\hline

				G1 & E1 & 11508 & 11406.33 & 11424 & 11624 &
				G22 & E1 & 13070 & 12922.83 & 12947 & 13359 \\

				& E2 & 11583 & 11528.67 & 11545 &  &
				& E2 & 13305 & 13226.14 & 13246 &  \\

				& E3 & 11613 & 11572.68 & 11589 &  &
				& E3 & 13348 & 13309.61 & 13325 &  \\

				& E4 & 11618 & 11582.66 & 11600 &  &
				& E4 & 13354 & 13324.92 & 13340 &  \\

				& DH & 11624 &  &  &  &
				& DH & 13357 &  &  &  \\
				\hline

				G6 & E1 & 2051 & 1962.33 & 1981 & 2178 &
				G23 & E1 & 13056 & 12917.78 & 12939 & 13344 \\

				& E2 & 2126 & 2074.80 & 2084 &  &
				& E2 & 13259 & 13204.25 & 13216 &  \\

				& E3 & 2155 & 2116.35 & 2124 &  &
				& E3 & 13327 & 13289.88 & 13298 &  \\

				& E4 & 2163 & 2127.00 & 2133 &  &
				& E4 & 13332 & 13306.46 & 13314 &  \\

				& DH & 2177 &  &  &  &
				& DH & 13339 &  &  &  \\
				\hline

				G11 & E1 & 434 & 374.04 & 382 & 564 &
				G36 & E1 & 7276 & 7195.30 & 7206 & 7680 \\

				& E2 & 550 & 529.37 & 534 &  &
				& E2 & 7609 & 7564.51 & 7570 &  \\

				& E3 & 560 & 553.30 & 554 &  &
				& E3 & 7665 & 7642.98 & 7646 &  \\

				& E4 & 562 & 555.43 & 558 &  &
				& E4 & 7672 & 7654.77 & 7658 &  \\

				& DH & 562 &  &  &  &
				& DH & 7672 &  &  &  \\
				\hline

				G14 & E1 & 2928 & 2869.56 & 2878 & 3064 &
				G39 & E1 & 2118 & 2020.11 & 2038 & 2408 \\

				& E2 & 3041 & 3020.89 & 3025 &  &
				& E2 & 2367 & 2323.35 & 2329 &  \\

				& E3 & 3061 & 3050.80 & 3053 &  &
				& E3 & 2406 & 2382.33 & 2386 &  \\

				& E4 & 3062 & 3055.47 & 3057 &  &
				& E4 & 2407 & 2389.98 & 2393 &  \\

				& DH & 3062 &  & &  &
				& DH & 2407 &  &  &  \\
				\hline

				G15 & E1 & 2909 & 2859.15 & 2867 & 3050 &
				&    &      &         &      & \\

				& E2 & 3030 & 3008.23 & 3012 &  &
				&    &      &         &      & \\

				& E3 & 3049 & 3038.46 & 3041 &  &
				&    &      &         &      & \\

				& E4 & 3050 & 3043.06 & 3045 &  &
				&    &      &         &      & \\

				& DH & 3050 &  &  &  &
				&    &      &         &      & \\

			\end{tabular}
		\end{ruledtabular}%
	}
\end{table*}

\begin{acknowledgments}
The authors thank Cemal Yalabık for useful discussions and Alex Kamenev
for a helpful question concerning the replica protocol. This work was supported by the TÜBİTAK 2232-A International Fellowship for
Outstanding Researchers Program (Project No. 124C530), under which
Ö. Ö. and A. C. K. received support. The numerical calculations reported
in this paper were performed at TÜBİTAK ULAKBİM, High Performance and Grid
Computing Center (TRUBA resources). All responsibility for the content of
the publication rests with the authors and is not scientifically
endorsed by TÜBİTAK. OpenAI ChatGPT (GPT-5.6 Sol) and OpenAI Codex (GPT-6 Sol) were used
during manuscript preparation to assist with literature synthesis,
inspection of code for potential bugs,  formal code
refactoring, source inspection, revision of scientific writing, and
mathematical consistency checks. All AI-assisted
outputs were critically reviewed by the authors; mathematical details,
cited literature, numerical values, scientific claims, and relevant code
changes were independently verified against the primary sources,
archived data, underlying code, and calculations. The scientific ideas,
methodology, and conclusions are those of the authors, who take full
responsibility for the content of the manuscript.
\end{acknowledgments}

\section*{Data and Code Availability}

The source code, manuscript source, numerical data archives underlying the
SK analyses, and the G-set benchmark inputs and retained checkpoints are
available in the
\href{https://github.com/omeronder2604/Statistical-Landscape-Exploration-and-Weak-Field-Defects-in-Selective-Oscillator-Ising-Machines}
{public GitHub repository}~\cite{OIMplusRepository}.
The repository also contains a machine-readable transcription of the G-set
benchmark results reported in this work.

\appendix
\section{Disorder generation, replica randomization, and reproducibility}
\label{app:randomization}

For each disorder realization, a single SK coupling matrix is generated and
shared by all 402 replicas throughout the oscillator evolution and the
subsequent diagnostic MH refinement. Independent Gaussian matrix entries are
first drawn with mean $J_0/N$ and variance $1/N$, after which the matrix is
symmetrized by averaging opposite off-diagonal elements and the diagonal is
set to zero. The resulting couplings therefore satisfy
\begin{equation}
J_{ij}\sim
\mathcal{N}\!\left(\frac{J_0}{N},\frac{1}{2N}\right),
\qquad i\neq j .
\end{equation}
The $J_0=0$ and $J_0=-4$ ensembles use the same centered disorder
realizations, with the latter differing by the corresponding uniform
off-diagonal mean shift. The same coupling matrix is also used for the
OIM-based and pure-MH calculations associated with a given disorder
realization.

The oscillator replicas are initialized once per realization with distinct
phases drawn uniformly from $[0,2\pi)$. During stochastic evolution, the
noise increment applied to each oscillator is
$K_n(t)\sqrt{\Delta t}\,z_i$, where the $z_i$ are standard-normal
pseudorandom variates. Noise draws are distinct across oscillator sites,
replicas, integration steps, and disorder realizations. At each selection
stage, cloning copies the oscillator phase configuration but not the random
number generator state. The descendant replicas therefore begin from the
same selected phase state but subsequently experience distinct stochastic
noise histories.

For the 30-realization SK OIM+/DH-OIM+ production runs, all stochastic
components are generated from a root seed of 42 using deterministic
\texttt{SeedSequence} branches for disorder generation, replica
initialization, oscillator epochs, individual replicas, and diagnostic MH
refinement. The representative single-realization calculation uses the same
seed-hierarchy construction with root seed 52. This deterministic seed
hierarchy makes reruns reproducible within a fixed implementation and
numerical environment while assigning distinct pseudorandom streams to
different replica histories. The archived SK OIM+/DH-OIM+ analysis files
retain the coupling matrices and paired pre- and post-refinement spin
configurations, but not oscillator phase histories, noise trajectories, RNG
states, explicit seeds, code hashes, or software versions.

\section{G-set Max-Cut benchmarks at intermediate epochs}
\label{app:Intermediate}

To complement the final G-set benchmark results in Table~\ref{tab:gset_results}, we examine how
the optimization progresses across the four oscillator epochs. Table~\ref{tab:gset_epoch_cuts}
reports the maximum and mean cut values after each epoch, together with the
cut threshold defining the best-performing third of the replica population.
Across the tested instances, the population shifts systematically toward
higher cut values from epoch 1 to epoch 4, with the largest improvements occurring in
the earlier epochs and progressively smaller gains as the annealing proceeds.
The subsequent short DH refinement provides a final local correction and
yields the DH--OIM+ values reported in Table~\ref{tab:gset_results}.

\bibliography{OIMp_v4}

\end{document}